\documentclass[pra, aps, twocolumn, superscriptaddress, nofootinbib, 10pt]{revtex4-2}

\usepackage{goldschmidt_aps}

\begin{document}

\title{Quantum Noise Suppression at Scale with Crosstalk-Robust Gate Sets}

\author{Andy J. Goldschmidt}
\email{Corresponding author: andyjgoldschmidt@gmail.com}
\affiliation{Department of Computer Science, University of Chicago, Chicago, IL 60637}

\author{Emilio Pel\'{a}ez Cisneros}
\affiliation{Department of Computer Science, University of Chicago, Chicago, IL 60637}

\author{Ryan Sitler}
\affiliation{Johns Hopkins University Applied Physics Laboratory, Laurel, Maryland 20723, USA}

\author{Kevin Olsson}
\affiliation{Johns Hopkins University Applied Physics Laboratory, Laurel, Maryland 20723, USA}

\author{Kaitlin N. Smith}
\affiliation{Department of Computer Science, Northwestern University, Evanston, IL 60208}

\author{Gregory Quiroz}
\affiliation{Johns Hopkins University Applied Physics Laboratory, Laurel, Maryland 20723, USA}
\affiliation{William H. Miller III Department of Physics \& Astronomy, Johns Hopkins University, Baltimore, Maryland 21218, USA}

\thispagestyle{plain}
\pagestyle{plain}

\begin{abstract}
We introduce crosstalk-robust gate sets, which are obtained using a novel, scalable optimal control problem exploiting locality.
Through the suppression of pairwise quantum crosstalk, the gate sets enable robustness that extends to multi-qubit circuits.
The IBM Quantum Platform devices provide a testbed for our gate sets, where we study their efficacy via error suppression protocols and randomized parallel single-qubit circuits of up to eight qubits.
Furthermore, we provide the first known assessment of the impact of complete optimal control gate sets on quantum algorithms.
Using a Hamiltonian simulation of a four-qubit transverse field Ising model, we show that noise-informed gates enhance median algorithmic performance by a factor of four over baseline Gaussian gates using the same calibration procedures. 
Lastly, we provide numerical evidence that optimized gate sets enable larger qubit-qubit coupling strengths that can cut two-qubit gate times in half. This result confirms that hardware-software co-design using quantum optimal control can create new opportunities for quantum computing architectures.
\end{abstract}

\maketitle

 \begin{figure}[t]
    \centering
    \includegraphics[width=\columnwidth]{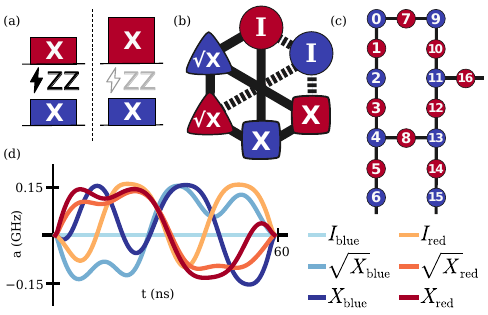}
    \caption{(a)~Identical $X$ gates are harmed by $ZZ$ crosstalk when played in parallel, but distinct (optimized) $X$ gates can operate unaffected. (b)~Optimal control finds crosstalk-robust gate sets by solving for gates (vertices) subject to $ZZ$ crosstalk errors (edges). Solid edges connect shaped gates, and dashed edges connect to the preserved idle identity. (c)~Two-coloring a heavy-hex lattice of qubits enables device-wide crosstalk mitigation. (d)~The pulse shapes of the gate set are shaped $X$ rotations equipped with standard calibration routines.
    }
    \label{fig:main-figure}
\end{figure}
\section{Introduction} \label{sec:introduction}
As quantum processors grow larger, correlated noise effects increase, and quantum crosstalk becomes as a key obstacle to scalability~\cite{gambetta2017building,mckay2019three,krinner2022realizing,mckay2023benchmarking}. For example, static crosstalk between neighboring qubits can inadvertantly entangle pairs undergoing supposedly single-qubit operations. Crosstalk degrades computational fidelity and is notably prevalent across a variety of quantum computing modalities like superconducting qubits~\cite{mckay2019three}, neutral atoms~\cite{graham2022multi}, ion traps~\cite{wright2019benchmarking}, and spin qubits~\cite{throckmorton2022crosstalk}.

Crosstalk error suppression has become an essential ingredient in demonstrations of progress in quantum hardware~\cite{wesdorp2026mitigating}. It has enabled successful demonstrations of quantum benchmark applications~\cite{arute2019quantum, kim2023evidence} in the pursuit of quantum advantage. Furthermore, crosstalk suppression has played an important role in the development---and beyond break-even demonstration---of quantum error detecting and correcting codes~\cite{parrado2021crosstalk,quiroz2024dfs,acharya2024quantum,vezvaee2025qed}. 

Work on methods for crosstalk suppression can be divided into hardware and software approaches. Hardware approaches include carefully-designed circuit elements~\cite{ku2020suppression, zhao2020high, kandala2021demonstration,zhao2021suppression}, cancellation tones~\cite{wei2022hamiltonian}, or tunable couplers~\cite{stehlik2021tunable}. Precisely designing circuit elements for crosstalk suppression is a nontrivial engineering task~\cite{zhang2022high}, while adding controls can introduce new error sources or increase the calibration overhead of the device.

Alternatively, software approaches trade hardware overhead for optimizations in other layers of the quantum computing stack. Previous gate designs focused on single-qubit error suppression show local crosstalk suppression during isolated operations, but not global crosstalk suppression during arbitrary circuits~\cite{watanabe2024zz,chen2026scalable}. Various software approaches require context-dependent optimizations that can burden other layers of the software stack. Examples include: crosstalk suppression with dynamical decoupling~\cite{tripathi2022suppression,zhou2023quantum,vezvaee2025qed}, pulse or scheduling co-optimization~\cite{murali2020software,xie2022suppressing}, and context-aware compiling~\cite{seif2024suppressing}. Higher-derivative variants of derivative
removal by adiabatic gate (DRAG) pulse families have been shown to be highly-effective at reducing microwave crosstalk, but these pulse families do not generally perform as effectively during simultaneous gates~\cite{hyyppa2024reducing, wesdorp2026mitigating}; our approach specifically covers this case and can be augmented by similar corrective tones. Tensor networks have been combined with optimal control to reduce the exponential scaling inherent to large-scale crosstalk suppression~\cite{le2026mitigating}; in contrast, our framework achieves linear scaling by capturing the local structure of crosstalk in the optimization.

We introduce a general framework for constructing \emph{crosstalk-robust gate sets} (CRGS)—collections of single-qubit gates that suppress pairwise crosstalk when executed concurrently across a quantum device. These gate sets are found through a coordinated quantum optimal control problem using the open-source \texttt{Piccolo.jl} framework~\cite{piccolo2025}, and can be integrated seamlessly into existing calibration and compilation flows. Our method, summarized in Fig.~\ref{fig:main-figure}, applies to arbitrary circuits and does not assume a specific hardware platform. 

Improved suppression of dominant errors at the gate level are expected to result in enhanced performance at the algorithmic level. This key assumption has driven quantum optimal control for many years~\cite{koch2022quantum}; however, the full use of quantum optimal control for error suppression at the scale of algorithms is rarely explored. We use Qiskit Pulse~\cite{alexander2020qiskit} access of IBM Quantum Platform processors to test the integration of our gate sets into quantum algorithms. To the best of our knowledge, this is the first known experimental assessment of the impact of optimal control on quantum algorithms. 

In our demonstrations, we calibrate and execute CRGS on IBM Brisbane, an Eagle processor with a fixed-coupler superconducting qubit architecture and a heavy-hex layout~\cite{gambetta2017building, hertzberg2021laser} and a similar in-house device with a ring layout and six qubits. A known challenge for this class of device is the emergence of parasitic $ZZ$ interactions directly caused by the always-on coupling. The coherence-limited fidelity of applications, measured by two-qubit gate speeds, is limited by the trade-off between the interaction strength and magnitude of $ZZ$ errors.

Using only standard calibration tools, we evaluate our gates with randomized single-qubit circuits and error-suppression protocols involving up to eight qubits, as well as a four-qubit simulation of the transverse-field Ising model (TFIM). Compared to unoptimized gates, our robust gates exhibit significantly reduced $ZZ$-type crosstalk, as measured across our multi-qubit benchmarks. In particular, we observe a $16\%$ $(1.016\times)$ improvement in gate fidelity when performing a random Clifford circuit simultaneously on four qubits. Furthermore, we find a $4\times$ median improvement in TFIM simulation performance. We also leverage purpose-built simulation tools~\cite{pulsesim2025} based on Qiskit Dynamics~\cite{puzzuoli2023qiskit} to explore stronger coupling regimes and show that our method supports higher gate speeds without performance loss, suggesting an opportunity for hardware-software co-design using quantum optimal control.

Our novel contributions are summarized as follows:
\begin{itemize}
    \item[$\circ$] We formulate crosstalk suppression as a scalable optimal control problem based on the connectivity graph of a quantum processor. Our algorithm enables construction of gate sets that globally suppress pairwise crosstalk errors.
    \item[$\circ$] We demonstrate straightforward integration of our gate sets into application workflows: no additional control lines, hardware modifications, or nonstandard calibrations are required. We experimentally validate our method on an IBM Eagle processor, achieving substantial crosstalk mitigation across multiple benchmark protocols and a quantum algorithm.
    \item[$\circ$] We present a study of the design tradeoffs between calibration efficiency and robustness, and demonstrate how our gates can co-design with hardware to push beyond current architectural limitations.
\end{itemize}

Taken together, our results show that optimized quantum control can reshape the design space for quantum hardware. By combining hardware-aware control with crosstalk-aware compilation, we achieve global $ZZ$ suppression in experiments and show a path to faster two-qubit gates without additional hardware complexity. Our work represents the first comprehensive assessment of the impact of pulse-level gate set design on algorithmic performance in realistic architectures. The framework of CRGS opens new paths for scalable quantum computing across diverse hardware platforms.

The manuscript is outlined as follows: In Sec~.\ref{sec:background}, we provide background information on gate sets, a Hamiltonian model for crosstalk, and our approach to quantum optimal control. Sec.~\ref{sec:crg} introduces a scalable, quantum optimal control protocol for designing CRGS that globally suppress arbitrary pairwise crosstalk errors that otherwise emerge when gates are played concurrently. Experiments evaluating CRGS---specifically designed for $ZZ$~crosstalk on IBM quantum processors---are discussed in Sec~\ref{sec:ibm-experiments}. In Sec.~\ref{sec:codesign}, we describe a co-design strategy for reducing two-qubit gate durations on IBM quantum processors, which is informed by CRGS. Finally, Sec.~\ref{sec:conclusion} summarizes our contributions and suggests directions for future work.

\section{Background} \label{sec:background}
 
\subsection{Gate sets} \label{sec:gs}
The \textit{Solovay–Kitaev theorem} states that arbitrary qubit rotations can be achieved with a fixed and finite set of single qubit gates~\cite{dawson2005solovay}. Commonly, these basis gate sets are chosen from the \textit{native} operations of the quantum processor. Native operations are rotation operators $U(\theta) = \exp(-i \theta H)$, about a hardware-accessible Hamiltonian, $H$. For example, control of a superconducting qubit can be modeled as $H(a(t)) = a(t) \sigma_X / 2$, where $a(t)$ is the envelope of a microwave drive that is in-phase and on-resonance with the qubit~\cite{krantz2019quantum}. In this case, the single-qubit native gates include $X$ rotations, $R_X(\theta) = \exp(-i \theta \sigma_X / 2)$ allowing for $X = \exp(-i \pi \sigma_X / 2)$ and $\sqrt{X} = \exp(-i \pi \sigma_X / 4)$.

One straightforward choice for the envelope, $a(t)$, is a truncated Gaussian, which includes the necessary rise and fall times for turning the drive on and off. Alternatively, more complicated control pulses can be used to change the dynamics of the qubit as it travels to its final state, $R_X(\theta) = \exp(-i \theta \sigma_X / 2) = \exp\ (-i \int_0^t a(t) dt\,  \sigma_X )$ with $\int_0^t a(t) dt = \theta$. This flexibility enables gates with useful additional properties, such as robustness to incoherent and coherent errors~\cite{khaneja2005optimal, carvalho2021error, propson2022robust, poggi2024universally, amer2025implementing}. 

\subsection{Crosstalk} \label{sec:computer}
As quantum computers scale to $10$s and $100$s of qubits, we see the emergence of correlated, non-local errors called \textit{crosstalk}~\cite{sarovar2020detecting}. Crosstalk can arise from unintended classical electromagnetic couplings--referred to as classical crosstalk--or quantum crosstalk, which comes from residual quantum couplings~\cite{patterson2019cr, wang2022control}. Both errors can be highly detrimental to device performance. Here, we will focus on addressing quantum crosstalk. A Hamiltonian model for pairwise quantum crosstalk errors is $H(t)=H_C(t) + H_{XT}$, where $H_C(t)=\sum_j a_j(t) \sigma^{(j)}_X$ models the otherwise perfect control of individual qubits. The error generating Hamiltonian is given by
\begin{equation} \label{eqn:qubit-crosstalk}
    H_{XT} = \sum_{\langle i,j \rangle} \sum_{\substack{A, B \in \\ \{X,Y,Z\}}}  \zeta^{(ij)}_{AB}\, \sigma^{(i)}_A \sigma^{(j)}_B,
\end{equation}
where $\zeta^{(ij)}_{AB}$ denotes the crosstalk coupling strength and $\sigma^{(j)}_A := I \otimes \dots \otimes \sigma_A \otimes \dots \otimes I$ acts on qubit $j$. This model captures experimentally-observed crosstalk effects across a wide variety of quantum computing systems.

\subsection{Quantum optimal control} \label{sec:qc}
Quantum optimal control (QOC) is a highly-successful framework for designing controls and managing the trade-offs among multiple competing objectives like gate fidelity, ease-of-calibration, and robustness to known error sources~\cite{khaneja2005optimal,carvalho2021error,propson2022robust,koch2022quantum}. \texttt{Piccolo.jl}~\cite{piccolo2025} is a state-of-the-art software package for solving QOC problems as constrained trajectory optimization problems; this formulation offers us a number of unique advantages~\cite{goldschmidt2022model,propson2022robust, trowbridge2023direct}. For our purpose, \texttt{Piccolo.jl} will enable us to introduce fidelity constraints, set bounds on control amplitudes and curvatures (i.e., bandwidths), and conveniently write down robustness objectives.

In trajectory optimization problems, both the quantum states and controls are optimization variables. Denote $\mathbf{X} = \begin{bmatrix}  \boldsymbol{x}_1 & \boldsymbol{x}_2 & \dots & \boldsymbol{x}_T \end{bmatrix}$ as the state matrix and $\mathbf{A} = \begin{bmatrix} \boldsymbol{\alpha}_1 &  \boldsymbol{\alpha}_2 & \dots &  \boldsymbol{\alpha}_T \end{bmatrix}$ as the actuation matrix, for state vectors $\boldsymbol{x}_t$ and control vectors $\boldsymbol{\alpha}_t$ across snapshot times,~$t$. These states and controls represent augmented states and controls~\cite{propson2022robust}; our state can include not only the (vectorized) unitary matrix, but also the control values $\mathbf{a}_t$ and control velocities $\dot{\mathbf{a}}_t$, so $\boldsymbol{x}_t = \begin{bmatrix} \vec{U}_t; & \mathbf{a}_t; & \dot{\mathbf{a}}_t \end{bmatrix}$. Similarly, the control vectors contain the accelerations of the controls and perhaps even the timesteps of the problem, $\boldsymbol{\alpha}_t = \begin{bmatrix} \ddot{\mathbf{a}}_t; & \Delta t_t \end{bmatrix}$. By using piecewise-constant acceleration as our control input, we are able to guarantee continuously differentiable pulse shapes without introducing further restrictions. Finally, both the state and the actuation matrices are collected into a trajectory matrix $\mathbf{Z} := \begin{bmatrix}  \boldsymbol{z}_1 & \boldsymbol{z}_2 & \dots & \boldsymbol{z}_T \end{bmatrix} = \begin{bmatrix}\mathbf{X}; & \mathbf{A} \end{bmatrix} $.

\subsubsection{Unitary smooth pulse problem template} \label{sec:qc:smooth}
The most basic primitive in \texttt{Piccolo.jl} is an optimal control problem for achieving a gate, $G$, using smooth controls; it is given by
\begin{align} \label{eqn:smooth-pulse-problem}
    \arg \min_{\mathbf{Z}}\quad & |1 - \mathcal{F}(\mathbf{Z}, G)| + \frac{1}{2} ||\vec{\mathbf{Z}}||^2_{\mathbf{R}} \\
    \nonumber \text{s.t.}\qquad & \mathbf{f}(\boldsymbol{z}_{t+1}, \boldsymbol{z}_{t}, t) = 0 \quad \forall\, t \\
    \nonumber & |\boldsymbol{z}_{t} | \le \boldsymbol{b}_t \quad \forall\, t.  
\end{align}
where $||\vec{\mathbf{Z}}||^2_\mathbf{R} := \vec{\mathbf{Z}}^T \mathbf{R} \vec{\mathbf{Z}}$ regularizes the problem. The objective is fidelity, $\mathcal{F}(\mathbf{Z}, G):= |\Tr \{U_T^\dagger G\}|$; notice how the function directly extracts the final unitary, $U_T$, from the trajectory, $\mathbf{Z}$ (there is no need for a forward integration of controls). The dynamics (contained in some given control Hamiltonian, $H(\mathbf{a})$) enter the problem as a constraint and are denoted implicitly by $\mathbf{f}(\mathbf{Z}_{t+1}, \mathbf{Z}_t, t) = U_{t+1} - \exp\{-i \Delta t_t H(\mathbf{a}_t)\} U_t = 0$. The unitary smooth pulse problem inputs are the goal unitary, $G$, the regularization matrix, $\mathbf{R}$, and the bounds, $\mathbf{B} = \begin{bmatrix} \boldsymbol{b}_1 & \boldsymbol{b}_2 & \cdots & \boldsymbol{b}_T \end{bmatrix}$. By default, we set all of the control bounds to $1.0$; in practice, this depends on the units of the controls. Moreover, we regularize on the integral of the control and its derivatives, usually as a diagonal matrix across all timesteps, so $||\vec{\mathbf{Z}}||^2_{\mathbf{R}} = r \sum_t \Delta t_t \left( \mathbf{a}^T_t \mathbf{a}_t + \dot{\mathbf{a}}^T_t \dot{\mathbf{a}}_t + \ddot{\mathbf{a}}^T_t \ddot{\mathbf{a}}_t \right)$ for a scalar regularization parameter, $r \approx 0.01$.  

\subsubsection{Unitary robustness problem template} \label{sec:qc:robust}
In robust control, we replace the gate fidelity with a measure of the sensitivity of infidelity to an error Hamiltonian,
\begin{equation} \label{eqn:sensitivity}
    \mathcal{R}(\mathbf{Z}; H_\text{error}) = \frac{1}{d} \Re\Tr{ \big(\frac{1}{T}  \sum\nolimits_{t=1}^T U_t^\dagger H_\text{error} U_t \big)^2},
\end{equation}
for Hilbert space dimension $d$. This \textit{error susceptibility} measures the net impact of the error Hamiltonian on fidelity as the system evolves under the applied controls~\cite{poggi2024universally}. Minimizing the error susceptibility means maximizing the robustness of the fidelity to the error Hamiltonian. A detailed calculation of the error susceptibility objective is discussed in Appendix~\ref{apdx:robustness}. To preserve the quality of our gate while maximizing robustness, we move the gate fidelity $\mathcal{F}$ to a constraint, and we provide a required fidelity $\mathcal{F}_0$ as a problem input. Standard approaches, e.g.~\cite{carvalho2021error,poggi2024universally}, add a robustness penalty alongside the infidelity in the objective, but this can introduce a trade off that depends on managing the penalty hyperparameter. The constrained optimization used by \texttt{Piccolo.jl} eliminates this issue,
\begin{align} \label{eqn:robustness-problem}
    \arg \min_{\mathbf{Z}}\quad & \mathcal{R}(\mathbf{Z}; H_\text{error}) + \frac{1}{2} ||\vec{\mathbf{Z}}||^2_{\mathbf{R}}  \\
    \nonumber \text{s.t.}\qquad 
    & |1 - \mathcal{F}(\mathbf{Z}, G)| \le 1 - \mathcal{F}_0 \\
    \nonumber & \mathbf{f}(\boldsymbol{z}_{t+1}, \boldsymbol{z}_{t}, t) = 0 \quad \forall\, t \\
    \nonumber & |\boldsymbol{z}_{t} | \le \boldsymbol{b}_t \quad \forall\, t.
\end{align}

A common application of Eq.~\eqref{eqn:robustness-problem} is the design of gates that are robust to small misalignments between the qubit frequency and the carrier frequency of the microwave control drive~\cite{carvalho2021error,watanabe2024zz}. Such detuning-robust single qubit gates are found by setting $H_\text{error} = \Delta\, \sigma_Z$ for detuning $\Delta$. They are useful for suppressing single qubit errors on real systems.

\begin{figure}[t]
    \centering
    \includegraphics[width=\columnwidth]{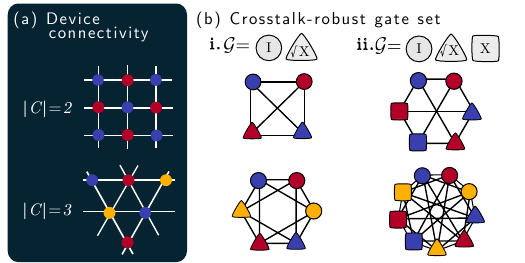}
    \caption{A crosstalk robust gateset (CRGS) is a single qubit gate set, $\mathcal{G}$, repeated in a set of colors, $\mathcal{C}$. (a) A two-colorable square lattice (top) and a three-colorable triangular lattice (bottom). (b) i. A two-element gateset and (b) ii. a three-element gateset are colored according to (a) and assembled into a graph $(V, E)$ for Eq.~\eqref{eqn:robust-gate set-problem}.}
    \label{fig:crg-example}
\end{figure}
\section{Designing crosstalk-robust gate sets} \label{sec:crg}
A \textit{crosstalk-robust gate set} (CRGS) is a single qubit gate set, $\mathcal{G}$, repeated in a set of colors, $\mathcal{C}$, with an orthogonality condition between gates of different colors. A CRGS can suppress all of the crosstalk for a given qubit layout if the number of colors is sufficient for a vertex coloring of that layout. In total, $\abs{\mathcal{G}} \cdot \abs{\mathcal{C}}$ colored gates are needed for a CRGS. For example, a square lattice of qubits is two-colorable, so two colors are needed in the CRGS. As a graph, a CRGS is a set of vertices and edges, $(V,E)$, where each vertex in $V$ is a colored gate, and edges in $E$ connect differently-colored gates according to the pairwise crosstalk errors that must be suppressed. Examples of this graph for different gate sets and layouts are shown in Fig.~\ref{fig:crg-example}.

In some cases it is possible to analytically find pairs of gates that are robust to specific pairwise crosstalk errors. For example, in Appendix~\ref{apdx:analytic-xtalk-robustness} we show that a pair of $X$ gates implemented using $2\pi$-offset square waves can guarantee zero $ZZ$ crosstalk susceptibility [Eq.~\eqref{eqn:sensitivity}] should those gates be played synchronously. In contrast, identical $X$ gates played in the same manner will experience the expected parasitic effects from the crosstalk. An illustration of this analysis was used for Fig.~\ref{fig:main-figure}(a). In general, we rely on optimal control problems like Eq.~\eqref{eqn:robustness-problem} to find more experimentally-viable pulse shapes.

\begin{figure}[t]
    \centering
    \includegraphics[width=\columnwidth]{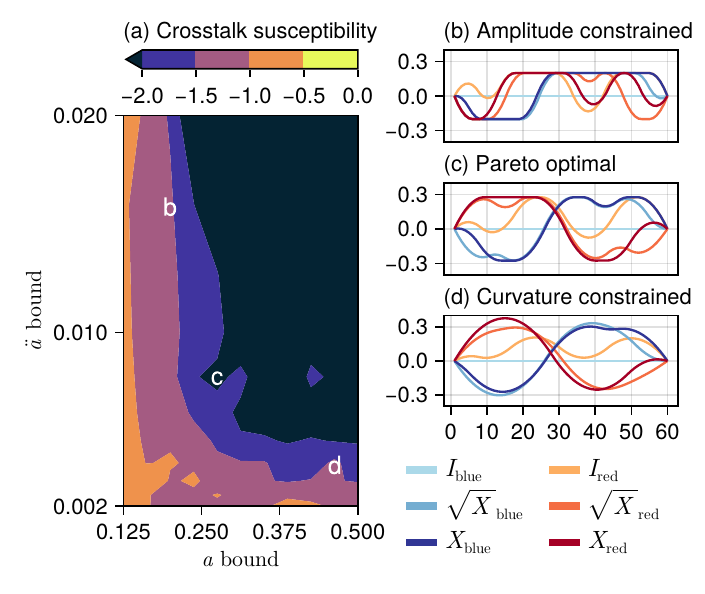}
    \caption{Crosstalk-robust gate sets optimized via Eq.~\eqref{eqn:robust-gate set-problem}. Each individual gate is constrained to have a minimum fidelity of $0.9999$. The log of crosstalk susceptibility is plotted over various amplitude and curvature bounds (note the suppressed zero on the $x$ and $y$ axis). Insets (b)-(d) show control amplitudes (GHz) and times (ns) for selected cases.
    }
    \label{fig:gate-set}
\end{figure}

\subsection{Crosstalk-robust gateset problem template} \label{sec:crgs-pt}
Naive application of Eq.~\eqref{eqn:robustness-problem} to the CRGS problem would require a Hilbert space involving $|V| = |\mathcal{G}| \cdot |\mathcal{C}|$ qubits. Alternatively, we can use problem structure to dramatically reduce the size of this coordinated optimal control problem. The \textit{crosstalk susceptibility}~\cite{zhou2023quantum} can be computed locally (see Appendix~\ref{apdx:xtalk-robustness}). That is to say, the total sensitivity [Eq.~\eqref{eqn:sensitivity}] to an arbitary set of pairwise crosstalk errors, $\sigma^{(i)}_A \sigma^{(j)}_B,\, \forall\, (i,j) \in E$, can be written as
\begin{equation} \label{eqn:xtalk-factor}
    \mathcal{R}\bigg(\mathbf{Z}; \sum_{i,j \in E} \sigma^{(i)}_A \sigma^{(j)}_B \bigg) = \sum_{i,j \in E} \bar{\mathcal{R}}\left(\mathbf{Z}_i, \mathbf{Z}_j; \sigma^{(i)}_A \sigma^{(j)}_B \right),
\end{equation}
where $\mathbf{Z}_j$ denotes the trajectory of qubit $j$ in the absence of crosstalk. Note that, critically, $\bar{\mathcal{R}}_{ij}$ depends on the crosstalk-free trajectory data $\mathbf{Z}_i$ and $\mathbf{Z}_j$ of qubits $i$ and $j$, not the full trajectory $\mathbf{Z}$. Hence, the CRGS problem template is 
\begin{align} \label{eqn:robust-gate set-problem}
    \underset{\mathbf{Z}_1, \mathbf{Z}_2, \dots, \mathbf{Z}_{|V|}}{\arg \min}\quad &  
    \sum_{i,j \in E} \bar{\mathcal{R}}\left(\mathbf{Z}_i, \mathbf{Z}_j; \sigma^{(i)}_A \sigma^{(j)}_B \right) \\
    \nonumber &+ \sum_{j \in V} \frac{1}{2} ||\vec{\mathbf{Z}}_j ||^2_{\mathbf{R}}  \\
    \nonumber \text{s.t.}\qquad 
    & |1 - \mathcal{F}(\mathbf{Z}_j, G_j)| \le 1 - \mathcal{F}_0 \quad \forall\, t\,, j \\
    \nonumber & \mathbf{f}(\boldsymbol{z}^{(j)}_{t+1}, \boldsymbol{z}^{(j)}_{t}, t) = 0  \quad \forall\, t\,, j \\
    \nonumber & |\boldsymbol{z}^{(j)}_{t} | \le \boldsymbol{b}^{(j)}_t \quad \forall\, t\,, j.
\end{align}
Contrast this formulation with the default crosstalk susceptibility, which is defined as a sum over all edges in $E$: in this case, the sensitivity objective [Eq.~\eqref{eqn:sensitivity}] involves each vertex in $V$, and scales exponentially. A CRGS problem template, through Eq.~\ref{eqn:xtalk-factor}, is an optimization over the independent trajectories of each qubit; hence, it scales linearly with the number of vertices. 

\subsection{Finding Pareto-optimal gate sets}  \label{sec:pareto}
Our optimal control scheme, represented by Eq.~\eqref{eqn:robust-gate set-problem}, allows us to efficiently explore crosstalk suppression as a function of the amplitude and curvature bounds on the controls, $\mathbf{a}$, and their accelerations, $\ddot{\mathbf{a}}$. In Fig.~\ref{fig:gate-set}, we plot the $ZZ$ crosstalk susceptibility, $\sigma^{(i)}_Z \sigma^{(j)}_Z$, of various implementations of a gate set involving $I$, $\sqrt{X}$, and $X$ gates. This is a complete gate set after including $Z$ rotations, which can be implemented entirely in software by tracking drive phases~\cite{mckay2017efficient}. We assume the device is two-colorable, and we impose an edge between $\text{red-}X$ and $\text{blue-}X$, $\text{red-}\sqrt{X}$ and $\text{blue-}\sqrt{X}$, and so on. 

The optimal crosstalk susceptibility defined in Eq.~\eqref{eqn:xtalk-factor}, summed over all edges $E$, is plotted as a function of the bounds on $\mathbf{a}$ and $\ddot{\mathbf{a}}$ in Fig.~\ref{fig:gate-set}(a). These results are area-constrained to achieve $ZZ$-robust controls without entering the regime of $2\pi$ offsets (e.g., the pulse implementing a $\text{red-}X$ and $\text{blue-}X$ accumulate the same rotation angle). This type of constraint will help later, during calibration with Qiskit Pulse; see Sec.~\ref{sec:calibration}. It is evident from Fig.~\ref{fig:gate-set} that larger amplitudes and curvatures are used for the CRGS. Indeed, more paths in the Hilbert space are available when the amplitude and curvature are large; eventually, the pulse shape is unrestricted enough that the tension binding fidelity and robustness loosens~\cite{poggi2024universally}. Under hardware constraints, the ability to perform a study such as this is critical to finding Pareto optimal pulse shapes that balance calibration ability and robustness.

\section{Demonstrations on Quantum Processors}

\begin{figure}[t]
    \centering
    \includegraphics[width=\columnwidth]{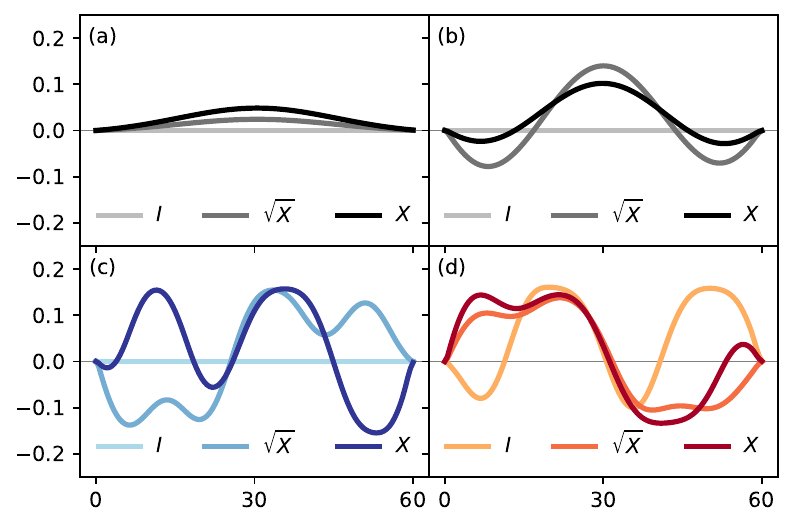}
    \caption{Gate sets generated and executed on IBM Brisbane. Pulse profiles shown for (a) Gaussian, (b) detuning-robust, and two-colorable, crosstalk-robust gate sets denoted as (c) blue and (d) red.}
    \label{fig:gates}
\end{figure}
\subsection{IBM Experiments: Background} \label{sec:ibm-experiments}
IBM's 127-qubit Eagle quantum processors use a heavy-hex layout of transmon qubits with static couplers and fixed qubit frequencies. These devices are known to experience coherent $ZZ$ crosstalk between adjacent qubits due to the always-on static coupling~\cite{mckay2019three, xie2022suppressing, zhou2023quantum, seif2024suppressing}. The perturbative size of the crosstalk between adjacent transmons $i$ and $j$ is
\begin{equation} \label{eqn:crosstalk}
    \zeta^{(ij)}_{ZZ}  = -J^2 \frac{2 (\alpha_i + \alpha_j)}{(\alpha_i + \Delta_{ij})(\alpha_j - \Delta_{ij})}
\end{equation}
where $\Delta_{ij} := \omega_i - \omega_j$ is the difference in transmon frequencies, $\alpha_j$ is the transmon anharmonicity, and $J$ is the fixed coupling strength~\cite{magesan2020effective}. Observe that this $ZZ$ interaction appears at second order in the coupling strength, $J$. Large coupling rapidly increases this crosstalk, which harms multi-qubit processes and makes scaling static coupler architectures challenging~\cite{zhao2022quantum, berke2022transmon}.

Qiskit Pulse~\cite{alexander2020qiskit} is used to construct and calibrate our crosstalk-robust gate on IBM Brisbane, an Eagle Processor. Subsequently, we evaluate the performance of the gate set through multi-qubit demonstrations. The IBM Brisbane device parameters are discussed in Appendix~\ref{apdx:simulation}. We fix our single qubit gate set to $I, \sqrt{X}, X$ and virtual $Z$ rotations, $R_Z(\phi)$~\cite{mckay2017efficient}, in accordance with the typical native gate set of these devices. We compare the performance of three classes of single qubit gate sets:
\begin{enumerate}
    \item[$\circ$] \textit{Gaussian} gate sets mimic the default shapes on IBM hardware.
    \item[$\circ$] \textit{Detuning-robust} gate sets, optimized using Eq.~\eqref{eqn:robustness-problem}, account for on-site $Z$ errors,  $H(a(t); \Delta) = \Delta\, \sigma_Z + a(t) \sigma_X / 2$. There is no robustness to $ZZ$.
    \item[$\circ$] \textit{Crosstalk-robust} gate sets (CRGS)  account for $ZZ$ crosstalk errors and on-site $Z$ errors. These are optimized using Eq.~\eqref{eqn:robust-gate set-problem} with two colors because of the heavy-hex layout of IBM devices.
\end{enumerate}

IBM's default single qubit gates are $60$~ns Gaussians. Our gate durations are scaled to $240$~ns; this is the minimum required for cloud-based calibration due to the truncation effects discussed next in Sec.~\ref{sec:calibration} and further in Appendix~\ref{apdx:calibration}. Our Gaussian gates are also scaled. When we use the unmodified IBM gates, we report these as the \textit{default} gate set. Our gate set envelopes are shown in Fig.~\ref{fig:gates}. Notice the control amplitude and curvature differences. Robust control involves spending control amplitude and curvature to traverse paths that are not geodesic, so studies like Sec.~\ref{sec:pareto} are critical for obtaining easy-to-calibrate controls that have the desired robustness properties.

\subsubsection{Compatibility with virtual~Z gates} \label{sec:virtualz}
Virtual~Z gates complete our gate set. These rotations are implemented by adjusting the phase $\phi$ of the control drive, which redefines the rotation axis of the controls: considering small timesteps $\Delta t$, the rotation operator $R_X(a_t \Delta t) \approx \text{exp}(-i a_t \Delta t\,  \sigma_X / 2)$ becomes $R_{X, \phi}(a_t \Delta t) \approx \text{exp}(-i a_t \Delta t (\cos(\phi) \sigma_X + \sin(\phi) \sigma_Y ) / 2)$. It is reasonable to expect that a phase change might alter the behavior of the $ZZ$ crosstalk sensitivity, as specified by Eq.~\eqref{eqn:sensitivity}; however, for controls restricted to the XY plane, the $ZZ$ crosstalk sensitivity is phase invariant. We can see this by taking the case of a single qubit. The adjoint action of $U_t := R_{X, \phi}(a_t \Delta t)$ on $\sigma_z$ is $U_t^\dagger \sigma_Z U_t = \cos(a_t \Delta t) Z + \sin(a_t \Delta t) Z_\perp(\phi)$ for $Z_\perp(\phi) = -\sin(\phi) \sigma_X + \cos(\phi)\sigma_Y$. Observing that $Z^2=I$ and $Z_\perp(\phi)^2=I$ have nonzero trace while mixed products have zero trace, we can reduce the sensitivity in Eq.~\eqref{eqn:sensitivity} to $\mathcal{R}(\mathbf{Z}; \sigma_Z) = \frac{1}{T^2} \sum\nolimits_{t,t'=1}^T \cos(a_t \Delta t) \cos(a_{t'} \Delta t) + \sin(a_t \Delta t) \sin(a_{t'} \Delta t)$, a phase-invariant quantity. In Appendix~\ref{apdx:analytic-xtalk-robustness}, we remark on how this readily extends to $ZZ$ crosstalk sensitivity.

\begin{figure}[t]
    \centering
    \includegraphics[width=\columnwidth]{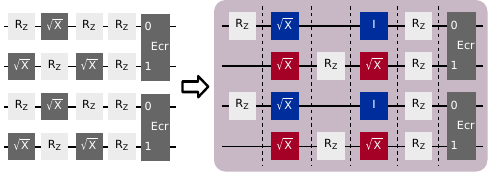}
    \caption{Illustration of a transpile pass to decompose the four-qubit circuit fragment into one-qubit, two-qubit, and virtual moments. The transpilation leverages the cancellation features afforded by the crosstalk-robust gate set to schedule overlapping single-qubit gates without accruing spatially-correlated errors.}
    \label{fig:transpile-pass}
\end{figure}
\subsubsection{Scheduling moments} \label{sec:scheduling}
To suppress $ZZ$ crosstalk, gates have to be played synchronously on adjacent qubits. Therefore, we designed a transpilation pass that takes any quantum circuit and decomposes the circuit into separate one-qubit, two-qubit, and virtual-$Z$ moments; an example is shown in Fig.~\ref{fig:transpile-pass}. As $Z$ rotations can be implemented virtually, we are only concerned with scheduling the single-qubit $X$ and $\sqrt{X}$ gates and the two-qubit echo cross-resonance (ECR) gate~\cite{sheldon2016procedure}. This is done in a way that maximizes the number of parallel gates in each moment. Accumulated virtual gates are applied at the start of each moment and are merged as needed. Crucially, the crosstalk-robust gate set affords flexibility in gate compilation. It is now possible to execute single-qubit gates simultaneously without accumulated crosstalk errors.

\subsubsection{Pulse calibrations} \label{sec:calibration}
To run our hardware experiments on the IBM Quantum Platform, we must first calibrate our pulses. Notably, the optimized gate sets do not require alternative calibration routines; they can use the exact same routines as the default pulse shapes. We discuss these routines (course calibration and fine calibration) in Appendix~\ref{apdx:calibration}. During calibration, high amplitudes led to a truncation effect that caused unrecoverable jumps in the coarse calibration curves. As a result, we increased the gate durations to fix the total rotation angle while preserving a low amplitude throughout the pulse. 

To guarantee good calibration results across all trials on IBM hardware, we chose to scale our pulses to $4 \times$ the duration of IBM's default $60$~ns pulses. We show that this $4 \times$ factor is artificial by using in-house transmon-based qubit hardware---which we refer to as the Applied Physics Laboratory (APL) system---to successfully calibrate our gate sets at durations consistent with IBM's default gates. Indeed, our RB experiments on in-house hardware (Section~\ref{sec:rb}) show that the calibration accuracy of our crosstalk-robust gate set does not diminish with decreasing gate duration. Ultimately, IBM hardware permits performance studies of our gate sets at scale, but in a different decoherence regime than the default gates.

When calibrating experiments involving up to eight qubits, we found a number of heuristics were useful for speeding up the runtime of our jobs. First, we used \textit{parallel calibration}---running the circuits for all qubits at once. 
Second, we used \textit{amplitude rescaling}---initializing the $X$ gate rough calibration value from the $\sqrt{X}$ gate. Calibration was successful whether or not these heuristics were applied; however, their use significantly lowered the QPU session resources required for cloud-based calibration. 

\begin{figure}[t]
    \centering
    \includegraphics[width=\columnwidth]{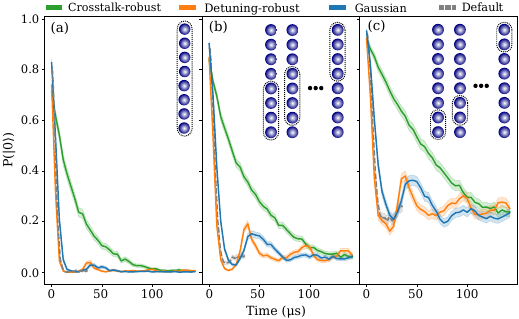}
    \caption{Ground state populations versus $XY4$ DD~sequence repetitions for (a)~the full eight qubit ground state population, (b)~an average over contiguous four-qubit combinations, and (c)~an average over contiguous two-qubit combinations. The duration of the default $XY4$ sequence is $4 \times$ shorter than the other cases. In all cases, the crosstalk-robust gate set significantly enhances decay rate over the alternative gate sets.}
    \label{fig:dd}
\end{figure}
\subsection{Dynamical decoupling experiment} \label{sec:dd}
%
\begin{table}[t]
    \centering
    \begin{tabular}{ccP{1.5cm}P{1.5cm}P{1.5cm}P{1.5cm}}
    \hline\hline
     & (MHz) & Crosstalk & Detuning & Gaussian & Default \\
    \hline
    \multirow{4}{*}{\ref{fig:dd}(a)} 
     & \(\gamma\)            & \textbf{0.0386}    & 0.198    & 0.131    & 0.163 \\
     & \(\pm\,\delta\gamma\) & 0.0004    & 0.008    & 0.006    & 0.009 \\
     & \(J\)                 & 0.0000    & 0.151    & 0.103    & 0.138 \\
     & \(\pm\,\delta J\)     & (fixed)   & 0.007    & 0.005    & 0.007 \\
    \hline
    \multirow{4}{*}{\ref{fig:dd}(b)} 
     & \(\gamma\)            & \textbf{0.0250}    & 0.107    & 0.093    & 0.124 \\
     & \(\pm\,\delta\gamma\) & 0.0005    & 0.009    & 0.008    & 0.006 \\
     & \(J\)                 & 0.0000    & 0.150    & 0.100    & 0.121 \\
     & \(\pm\,\delta J\)     & (fixed)   & 0.007    & 0.007    & 0.004 \\
    \hline
    \multirow{4}{*}{\ref{fig:dd}(c)} 
     & \(\gamma\)            & \textbf{0.0170}    & 0.083    & 0.09    & 0.106 \\
     & \(\pm\,\delta\gamma\) & 0.0008    & 0.008    & 0.02    & 0.006 \\
     & \(J\)                 & 0.0000    & 0.149    & 0.08    & 0.119 \\
     & \(\pm\,\delta J\)     & (fixed)   & 0.006    & 0.01    & 0.004 \\
    \hline\hline
    \end{tabular}
    \caption{Decay parameter \(\gamma\) and oscillation parameter \(J\) fit to $f(t)=(1+e^{-\gamma t} \cos(J t)) / 2$ for the $XY4$ experiments in Fig.~\ref{fig:dd}. Crosstalk-robust gates provide a $4$-$5\times$ improvement in the decay parameter (bold).}
    \label{tab:fit_gamma_J}
\end{table}

Dynamical decoupling (DD) sequences suppress single-qubit error channels like detuning by refocusing errors~\cite{viola98dynamical, viola1999dynamical, zanardi1999symmetrizing}. Parallel DD sequences are unable to refocus $ZZ$ crosstalk errors because of the even contributions from the parallel gates~\cite{zhou2023quantum}. We apply parallel DD sequences of the $XY4$ variety, defined by the circuit $XYXY$, in order to directly probe the strength of $ZZ$ crosstalk between qubits. This is accomplished by first preparing $N$ qubits in the state $\ket{\psi(0)}=|+\rangle^{\otimes N}$ using simultaneous unitaries $U_{\rm prep}=\otimes^{N}_{j=1} U_j$. We then subject the qubits to $n$ repetitions of parallel DD and then apply $U^\dagger_{\rm prep}$ prior to measuring in the computational basis. Because $XY4$ acts like a robust idling period, the circuit would ideally return the qubits to the ground state $\ket{0\cdots 0}$.

In our experiments on IBM Brisbane, we performed up to $n=200$ repetitions of the $XY4$ sequence on eight adjacent qubits for each of the gate sets introduced at the start of this section (see \ref{sec:ibm-experiments}). The results are shown in Fig.~\ref{fig:dd} using 2048 shots. Error bars are computed using the normal approximation to the binomial distribution. We fit our data to $f(t)=(1+e^{-\gamma t} \cos(J t)) / 2$ and report the results in Table~\ref{tab:fit_gamma_J}.

At sufficient sequence depth under $XY4$ (using non-crosstalk-robust gates), we observe an oscillation in the ground state population due to a precession caused by the always-on $ZZ$. We estimate this frequency to be about $0.2$~MHz (indeed, we use this to validate our simulation parameters in Appendix~\ref{fig:xy4-fit}). Observe that the crosstalk-robust gate set eliminates the oscillations that are otherwise present for Gaussian, default, and detuning-robust gate sets; thus, we see a clear signature of crosstalk cancellation. Our novel gate set provides a dramatic boost to fidelity, affording a decay parameter that is 30\% of that found for Gaussian pulses for the full eight-qubit demonstration.

We also show partial traces over smaller subsets of qubits in order to witness the oscillations coming from the $ZZ$. Notice that the Gaussian and detuning-robust oscillations essentially overlap because the frequency is determined by the $ZZ$ strength, and the $ZZ$ is completely unmitigated in both of these cases. In our experiments, we also report the default IBM gate set. It is important to note that because the default gate set is $4\times$ shorter than the other gate sets used in these experiments, there are $4\times$ as many default gates in the same duration. Nevertheless, the crosstalk-robust gate set yields significant improvements in the decay parameter over all alternative gate profiles considered.

\begin{figure}[t]
    \centering
    \includegraphics[width=\columnwidth]{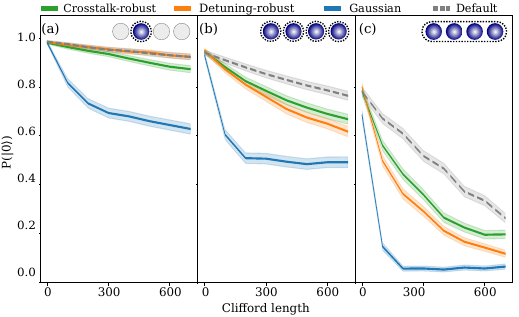}
    \caption{The probability of measuring the ground state after a string of random Clifford (RC) gates that simplify to the identity. (a)~RC applied to an individual qubit, (b)~RC applied simultaneously, showing the average over the ground state population of each qubit, (c)~RC applied simultaneously, showing the four-qubit ground state population. The simultaneous cases [panels (b),~(c)] inject crosstalk; in these cases, the crosstalk robust gate set outperforms all alternatives of comparable duration.}
    \label{fig:rb}
\end{figure}
\subsection{Random circuit experiments} \label{sec:rc}
\begin{table}[t]
    \centering
    \begin{tabular}{lP{0.66cm}P{1.4cm}P{1.4cm}P{1.4cm}P{1.4cm}}
    \hline\hline
     &  & Crosstalk & Detuning & Gaussian & Default \\
    \hline
    \multirow{2}{*}{$\,\,\,$\ref{fig:rb}(a)} 
     & \(p\) & 0.9998 & 0.9993 & 0.9939 & 0.9992 \\
     & \(\pm\) & 0.0002 & 0.0003 & 0.0005 & 0.0002 \\
     \hline
     \multirow{2}{*}{$\,\,\,$\ref{fig:rb}(b)} 
     & \(p\) & 0.9981 & 0.9982 & 0.9864 & 0.9993 \\
     & \(\pm\) & 0.0001 & 0.0001 & 0.0007 & 0.0001 \\
     \hline
     \multirow{2}{*}{$T_1$, $T^\text{echo}_2$} 
     & \(p_\text{lim.} \) & 0.9986 & 0.9986 & 0.9964 & 0.9996 \\
     & \(\pm \) & 0.0006 & 0.0006 & 0.0014 & 0.0001 \\
     \hline
     \multirow{2}{*}{$T_1$, $2T_1$} 
     & \(p_\text{lim.} \) & 0.9997 & 0.9997 & 0.9995 & 0.9999 \\
     & \(\pm \) & 0.0001 & 0.0001 & 0.0001 & 0.0001 \\
    \hline\hline
    \end{tabular}
    \caption{ The average single-qubit decay parameters from Fig.~\ref{fig:rb}(a) and~(b) [fit: $p(m) = a p^m + b$] for each gate set under the same random Clifford circuit (top), are shown alongside the decoherence-limited decay parameter computed from the gate duration and the reported $T_1$ and $T_2 \le 2 T_1$ at the time of calibration (bottom). The first two rows convey results for Fig. 7(a) and (b) using the various gate sets and the above fitting equation. The last two rows include estimates based on the average coherence limit from $T_1$ and $T_2$ values. Overall, a performance improvement is observed for shaped controls (crosstalk, detuning) over Gaussian. Default gates are $4 \times$ shorter, as they are not limited by the artifacts of cloud-based calibration (Sec.~\ref{sec:calibration}). Fig.~\ref{fig:rb-apl} provides evidence that this duration difference is artificial.
    }
    \label{tab:dep_alpha_fits}
\end{table}

To further assess the impact of our gate sets on gate fidelity, we play parallel random Clifford (RC) circuits on a four qubit chain from IBM Brisbane. The randomization means we no longer have the idling behavior with clear $ZZ$ oscillation that we observed in DD (Sec.~\ref{sec:dd}); however, we show that our protocol still benefits from shaped controls, which help overcome drive-induced dephasing noise like the state-dependent injections from always-on $ZZ$. We generate a one-qubit RC sequence such that the final unitary is the identity, and use the same RC circuit for each of the four qubits. Running the same circuit during the simultaneous protocol ensures the impact of crosstalk.

The results of our experiments are shown in Fig.~\ref{fig:rb}. From left to right, we show an isolated (one qubit) RC experiment, then two versions of a simultaneous (four qubit) RC experiment: in the middle, we have the average probability of a single qubit returning to the ground state, and on the right, we have the probability of the four-qubit chain returning to the ground state. In each, error bars are computed using the normal approximation to the binomial distribution. Observe that the CRGS maintains the best overall performance when going from isolated to simultaneous experiments (comparing the left and middle plots). Notice that for the individual experiments, the detuning-robust gate sets maintained a higher fidelity than the CRGS~[panel~\ref{fig:rb}(a)]. As such, it is more pronounced that the CRGS overtakes the performance of the detuning-robust gate set in the simultaneous experiment~[panel~\ref{fig:rb}(b)]. 

In Table~\ref{tab:dep_alpha_fits}, we report fits of our RC survival probability to the model $f(m)=a p^m + b$, where $p$ is the decay parameter, $m$ is the number of Clifford gates, and $a$ and $b$ are fit coefficients. For reference purposes, we provide the average coherence limited error of our four qubit chain using IBM Brisbane's reported $T_1$ and $T_2$ at the time of each gate set's calibration. This limit can be determined via $\epsilon = (3 - 2e^{-t_d/T_2} - e^{-t_d/T_1})/6$, where $t_d$ is the gate duration (the corresponding decay parameter is $p_\text{lim.} = 1- 2\epsilon$)~\cite{gambetta2012characterization,magesan2012characterizing}. We also include the $T_1$-limited error, obtained by setting pure dephasing to zero, such that $T_2 = 2 T_1$ (Appendix~\ref{apdx:simulation}). 

By comparing relative RC performance via decay parameters for a fixed RC circuit, Table~\ref{tab:dep_alpha_fits} indicates that shaped pulses are able to suppress drive-induced errors in both the isolated and simultaneous cases. Conclusions rely on relative changes among gate sets. In the standard case, $T_2 = T_2^\text{echo}$ is characterized by a spin echo, which refocuses static detuning errors. The shaped controls (crosstalk, detuning) exceed the $T_2^\text{echo}$ limit and push the $2 T_1$ bound, indicating shaped controls overcome detuning errors that otherwise limit Gaussian controls~[rows 7(a)]. In part, exceeding this limit can be explained by differences between the time when $T_2$ values were recorded and the experiments were executed. Shaped controls also maintain their performance over Gaussians when additional spectator errors are injected during the simultaneous protocol~[rows 7(b)]. Default gates still maintain higher estimated probability due to their $4\times$ shorter duration. Furthermore, they do not appear to be limited by the challenges faced with cloud-based calibration. We show in Sec.~\ref{sec:rb} that this difference in performance is artificial and can be overcome with more transparent low-level access.

With tens more additional random circuits per Clifford length, we could naturally extend our results to standard (parallel) randomized benchmarking (RB)~\cite{knill2008rb, gambetta2012characterization}. In this limit, we anticipate some reconciliation of the estimated probabilities shown in the first two rows of Table~\ref{tab:dep_alpha_fits} with the coherence limit. This was not possible at the time of experimental execution due to eventual restricted hardware access by the provider. As an alternative to pursuing such statistics on IBM cloud-based hardware, we make use of the in-house APL system in the next section.

\begin{figure}[t]
    \centering
    \includegraphics[width=\columnwidth]{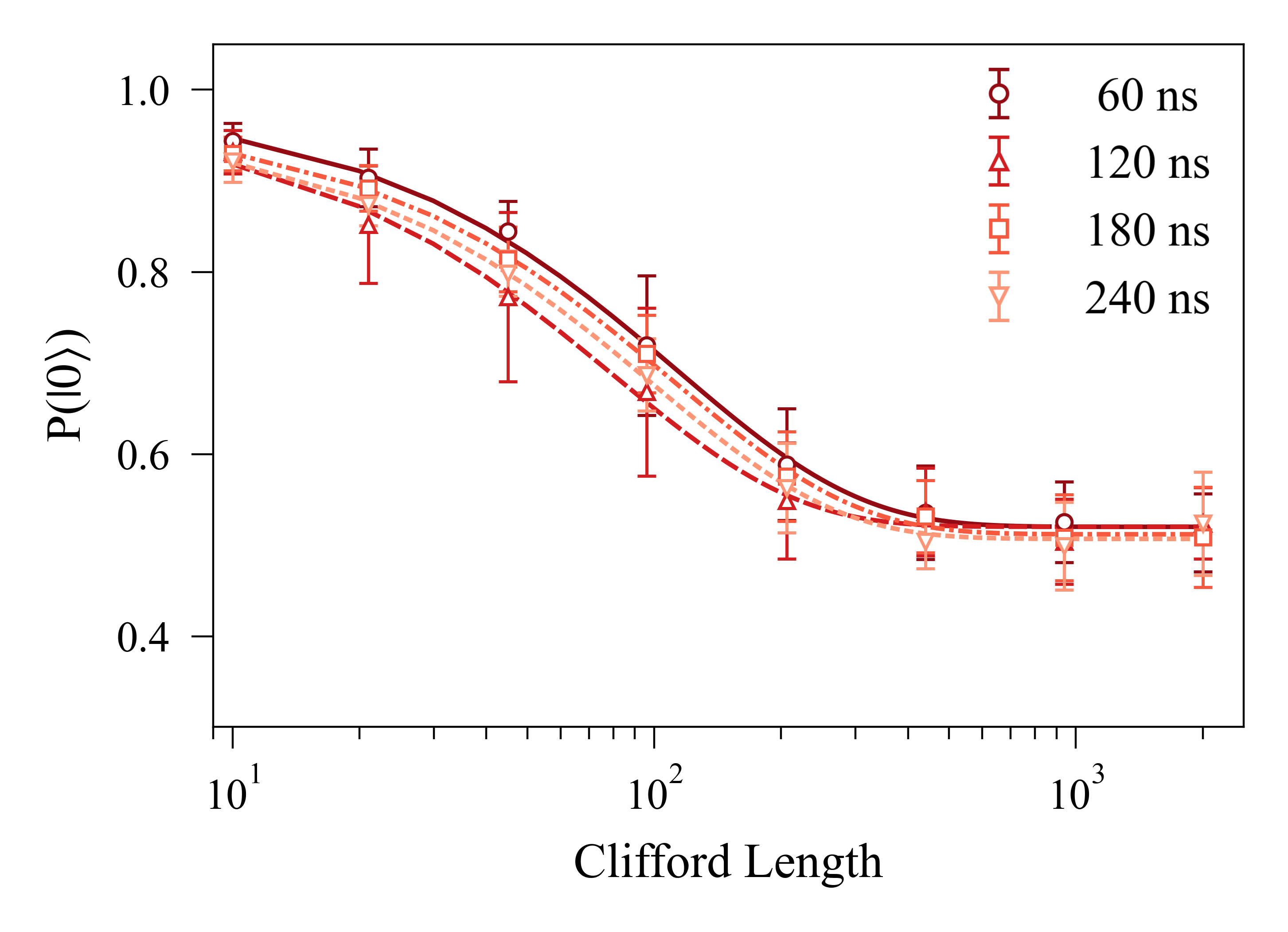}
    \caption{RB experiments performed on the APL system using optimized gates with various gate durations. RB decay curves are produced from 20 RB sequences per Clifford length. Data points and error bars denote average probability and standard deviation, respectively. We observe some deviation between different pulse durations, however, the probabilities are equivalent within error bars.}
    \label{fig:rb-apl}
\end{figure}

\subsection{Randomized benchmarking experiments} \label{sec:rb}
The in-house APL system allows us to directly compare gate set performance at default $60$~ns pulses using full RB experiments; information regarding the APL system can be found in Appendix~\ref{apdx:apl-system}. We use the APL system to reify that the $4 \times$ duration used in our IBM experiments is an artifact of the available IBM access, and not a physical limit of the CRGS themselves. The APL system did not readily permit four qubit operations due to overwhelming classical crosstalk challenges associated with the particular device used for this study; therefore, we use the APL system to complement the IBM Brisbane results. We performed single qubit RB experiments for CRGS at $1 \times$, $2 \times$, $3 \times$, and $4 \times$ the default duration using the APL system; see Fig.~\ref{fig:rb-apl} for the resulting RB curves. Each data point corresponds to average probability, with error bars denoting standard deviation. The statistics are calculated using 20 RB sequences per Clifford length.

The sensitivity of gate performance to pulse duration is assessed via the error per Clifford (EPC) and compared to the coherence-limited error rate; coherence times are given in Appendix~\ref{apdx:apl-system}. We obtain the following results for each gate duration: (1) 60 ns: EPC = $8.8\pm0.6\times 10^{-3}$, (2) 120 ns: EPC = $12\pm 1\times 10^{-3}$, (3) 180 ns: EPC = $9.0\pm 0.5\times 10^{-3}$, (4) 240 ns: EPC = $9.9\pm 0.6\times 10^{-3}$. The corresponding decoherence limits are (1) $0.7 \times 10^{-3}$, (2) $1.4 \times 10^{-3}$, (3) $2.1 \times 10^{-3}$, (4) $2.8 \times 10^{-3}$. We observe relatively consistent EPCs (within error bars) among the gate durations. Because the APL hardware and the IBM hardware are comparable architectures---both are fixed frequency, fixed coupler devices---these results suggest that shorter CRGS durations on IBM hardware should not result in reduced performance.

\begin{figure}[t]
    \centering
    \includegraphics[width=\columnwidth]{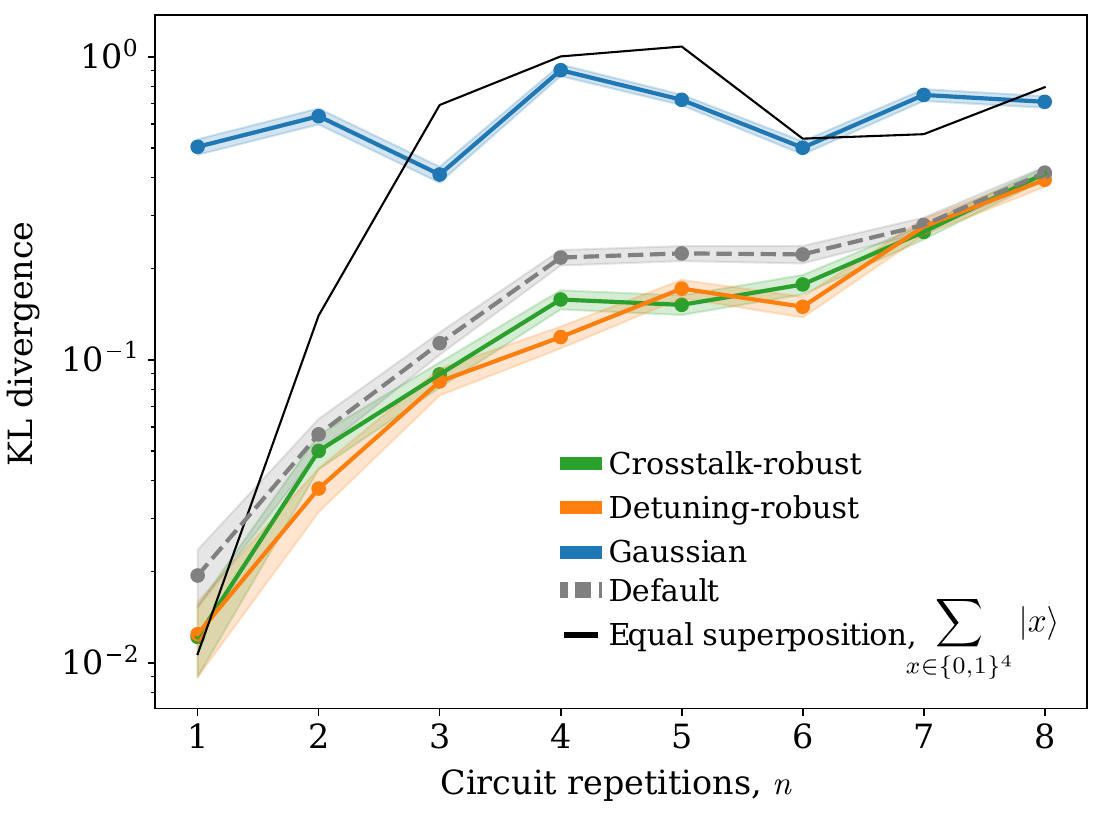}
    \caption{Kullback–Leibler (KL) divergence, comparing the experimental distributions to the ideal noise-free distribution, for repetitions of a transverse-field Ising model (TFIM) circuit. Successful implementations of the TFIM circuit remain below the equal superposition line. Demonstrations were performed on IBM~Brisbane using $n=4$ qubits and 2048 shots.}
    \label{fig:tfim}
\end{figure}
\subsection{Hamiltonian simulation demonstration} 
\label{sec:tfim}
We further assess the efficacy of our gate set by examining its impact on the performance of a quantum algorithm. Specifically, we consider the problem of Hamiltonian simulation~\cite{feynman2018simulating, georgescu2014sim} and utilize an IBM quantum processor to compare our various gate sets. On digital quantum computers, Hamiltonian simulation is performed via a time evolution operator that is digitally decomposed into a set of unitaries that approximately simulate the desired evolution. These unitaries are further compiled down to the native gate set of the hardware, i.e., the single- and two-qubit gate operations. We employ our optimized gate set and scheduling protocol at this layer to address potential errors in the Hamiltonian simulation generated by $ZZ$ crosstalk and detuning errors. A comparison is then made against non-optimized gate sets to determine if improvements at the physical (i.e. pulse) layer effectively propagate to the algorithmic layer.

We choose to simulate the transverse-field Ising model (TFIM) Hamiltonian, $H = g \sum_{\langle i, j \rangle} \sigma^{(i)}_x \sigma^{(j)}_x + h \sum_{j} \sigma^{(j)}_z $, where $g$ denotes the coupling strength and $h$ is the local bias on each spin. The dynamics generated by $H$ after a time $T$ are given by $U(T)=e^{-i H T}$. This time evolution operator can be approximately digitally simulated via the second-order product formula~\cite{suzuki1991general, berry2007efficient}
\begin{equation}
    U(T) \approx \left( 
    \prod_{i,j} U^{xx}_{ij}\left(\tfrac{\Delta t}{2}\right) 
    \prod_{k} U^{z}_{k}(\Delta t) 
    \prod_{i,j} U^{xx}_{ij}\left(\tfrac{\Delta t}{2}\right) 
    \right)^n.
    \label{eq:trotter}
\end{equation}
$T=n \Delta t$~represents the total time set by $n$ repetitions of the constituent unitaries applied for a time $\Delta t$. The decomposition above is described by two unitaries: $U^{xx}_{ij}(t)=e^{-i g t \,\sigma^{(i)}_x \sigma^{(j)}_x}$ and $U^{z}_{k}(t)=e^{-i h t \,\sigma^{(k)}_z}$. The former can be decomposed into a single qubit $X$-rotation on the $i$th qubit bookended by a CNOT gate from qubit $i$ to $j$ on each side. Compilation of the $X$-rotation into the IBM basis gate set permits the use of our optimized $\sqrt{X}$ gates, while the CNOT operation is decomposed into ECR and $\sqrt{X}$ gates. The remaining single qubit gates described by $U^{z}_{k}(t)$ correspond to virtual-$Z$ rotations.

We experimentally simulate the TFIM at its phase transition, $g = h = 2\pi$, with a time step $\Delta t = 0.05$. A system of four qubits is prepared in the equal superposition $\ket{\psi(0)}=\ket{++++}$ and subject to $n$ repetitions in accordance with Eq.~(\ref{eq:trotter}). We assess performance by comparing the measured output distribution of the device to that of the ideal, noise-free distribution for a range of circuit simulation times via the Kullback–Leibler (KL) divergence. This evaluation approach enables a $Z$-basis, observable-independent assessment of the hardware's performance when subject to our control protocols.

The results of the TFIM comparison are shown in Fig.~\ref{fig:tfim}. KL divergence is calculated for different control scenarios as a function of circuit repetitions, $n$. Data is collected from IBM~Brisbane using 2048 shots per repetition. Error bars on the KL statistic are computed using $1000$ bootstrapping iterations (sampling our total shots with replacement and recomputing the KL to estimate the population variance)~\cite{stine1989introduction}.

To provide a baseline, we plot the KL divergence of the equal-superposition state. This provides two useful reference points: one at the start ($n=1$) and one at the end ($n=8$). At the start, upon completing a single iteration of the TFIM simulation, our quantum state has not evolved far from its initial state, so the KL should be close to the value reported for the equal-superposition state. In order to interpret the benchmark endpoint, we first observe that in a $Z$-basis measurement, the equal superposition state and the totally-mixed state share a distribution. Therefore, we expect the KL divergences of all experiments to eventually saturate at the KL divergence of the equal superposition / totally-mixed state. Any experimental KL value below the equal superposition curve can be considered good, while any value saturating this curve can be considered bad. Observe that the divergence of the Gaussian gate set quickly saturates at the KL of the totally-mixed state. In contrast, we see that the detuning- and CRGS are $50 \times$ more accurate after a single Trotter step. The average performance boost is approximately $4\times$, and the enhanced performance persists significantly longer than the default gate set before converging to the totally-mixed state. 

Another important observation in this experiment is that the default gate set behaves approximately the same as the robust gate sets, even though we have seen from previous experiments that the default gates possess better coherence-limited fidelity. This is because the TFIM includes two-qubit gates, and these are the same for all of our gate sets. Two-qubit gates dominate the circuit runtime and set the coherence limit. It stands to reason that one way to get a better coherence-limited fidelity for our circuits is to make two-qubit gates faster. This requires increasing the coupling strength. A linear increase in coupling strength leads to a quadratic increase to crosstalk [Eq.~\eqref{eqn:crosstalk}],  so this presents an opportunity to pursue hardware-software co-design of CRGS and coupling strengths. A co-design strategy, which we study in the next section, can allow us to pursue circuit performance exceeding the current coherence-limited fidelity set by two-qubit gates.

\begin{figure}[t]
    \centering
    \includegraphics[width=\columnwidth]{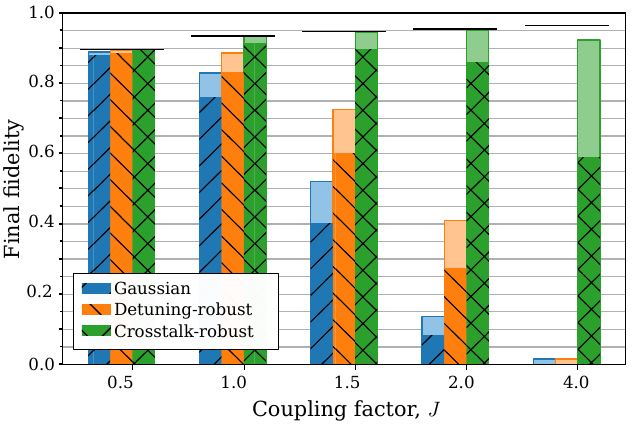}
    \caption{A gate set comparison of the simulated TFIM circuit fidelity at different coupling strengths, $J$, and $J=1$ corresponds to IBM Brisbane. The horizontal lines over the sets of bars indicate the decoherence-limited fidelity, which increases from left to right because two qubit gates are faster at larger $J$. The bars without patterns assume crosstalk-robust two-qubit gates. Any remaining gap (e.g., $J=4$) indicates that crosstalk has overwhelmed the suppression protocols.}
    \label{fig:bars}
\end{figure}
\section{Hardware-software co-design of gate sets and coupling strengths} \label{sec:codesign}
IBM's Eagle processors are the product of careful engineering that balances the trade-offs among device parameters. Nevertheless, one can open up this design space further by considering hardware-software co-design~\cite{murali2020software,li2021codesign,xie2022suppressing,seif2024suppressing,koch2022quantum,watanabe2024zz,smith2022scaling,lin2024codesign}. CRGS mitigate the crosstalk susceptibility of parallel operations, so larger hardware couplings $J$ can be tolerated by making a control software change. The benefit of large couplings is faster two-qubit gates, and therefore, less decoherence. In this section, we use simulations of our TFIM demonstration from Sec.~\ref{sec:tfim} to see how our gate sets might influence the choice of coupling strength.

Pulse-level simulations allow us to study crosstalk suppression outside the parameter regimes of the IBM hardware on which we run our demonstrations. In particular, we can explore increasing or decreasing the coupling strength, $J$, between statically coupled transmons. The two error sources modeled by our simulation---decoherence and crosstalk---are set by $J$. First, decoherence errors increase as $1/J$. This is because decoherence errors increase with duration, circuit duration is dominated by two-qubit gates, and two-qubit gate times scale with $1/J$. We model decoherence using a Lindblad master equation with a relaxation rate set by $T_1$ and a dephasing rate set by $T_2$~\cite{krantz2019quantum}. Second, $ZZ$ crosstalk errors increase as~$J^2$, which comes directly from Eq.~\eqref{eqn:crosstalk}. We set a baseline coupling strength $J$ such that $ZZ$ is approximately $200$~kHz for all edges. Full details of our simulation can be found in Appendix~\ref{apdx:simulation}.

In Fig.~\ref{fig:bars}, we show the final unitary fidelity after simulating a single repetition of the TFIM circuit for different factors of the IBM Brisbane coupling strength. We see that co-design of stronger coupling devices with CRGS can enable higher-fidelity static coupler devices for practical applications. Fig.~\ref{fig:bars} shows $J \times \{0.5, 1.0, 1.5, 2.0, 4.0\}$, simulating the same TFIM under $ZZ$ error and decoherence. Detuning error is not included to isolate the impact of crosstalk. Moreover, it was observed in the experiments of Sec.~\ref{sec:tfim} that detuning only harms the fidelity of the Gaussian case. Detuning is also independent of the coupling strength and would apply uniformly to each case. 

Using default gates, crosstalk errors start to dominate after even a $50\%$ increase in $J$, indicating that IBM devices are operating at an optimal point for the trade-off between gate speed and crosstalk error when Gaussian gates are used. However, we see that the ceiling of attainable fidelity is lifted by increasing the coupling strength, which comes from the reduced decoherence due to faster two-qubit gate times. A $2-3\%$ improvement in the fidelity ceiling is made possible by increasing $J=1$ to $J=2$ or $J=4$, respectively. 

In order to take advantage of this increase in the coherence limit, we must leverage the CRGS. This is evident from the patterned bars in Fig.~\ref{fig:bars}, where we observe a significant improvement enhancement over crosstalk-susceptible alternatives. For example, in the case of $J=2$, we find a $3.4\times$ and $11.3\times$ improvement in fidelity for the CRGS over detuning-robust and Gaussian pulse profiles, respectively.

Further improvements in fidelity can be made by enforcing crosstalk-robustness in two-qubit gates as well.
In particular, we can keep the gate durations the same, but modify the cross-resonance envelope to suppress $ZZ$ crosstalk. We use a detuning-robust drive to accomplish this (Appendix~\ref{apdx:simulation}). In Fig.~\ref{fig:bars}, the CRGS with crosstalk-robust two-qubit gates is denoted by the non-patterned bars. Note that with a $2\times$ increase in $J$, the addition of the robust two-qubit gate results in fidelities that are nearly at the coherence limit, approximately $1.11\times$ better than the single-qubit robust gate set, alone. Any remaining gap between the non-patterned bars and the decoherence limit indicates that the robust single-qubit and robust two-qubit protocols were insufficient to suppress the crosstalk contribution (e.g., $J=4$ in Fig.~\ref{fig:bars}).

\section{Conclusion} \label{sec:conclusion}
Crosstalk-robust gate sets (CRGS) demonstrate that optimal control can be used to globally suppress quantum crosstalk on quantum devices with minimal overhead: no additional control lines, hardware modifications, or nonstandard calibrations are required.
We convey this result through multi-qubit state preservation, gate characterization, and algorithmic performance in real quantum architectures. These demonstrations provided clear evidence that optimal control gate sets are a valuable design component in the quantum computing stack.

The gate set is designed via an optimal control problem that constructs crosstalk and detuning robust pulse profiles. We showed that this coordinated optimal control problem can be recast as a graph coloring problem informed by the topology of the hardware. Crucially, this perspective enables the construction of an orthonormal set of pulse profiles that scales linearly in the number of colors and native gate operations. Thus, the size of the control problem does not scale with the number of qubits, but rather the connectivity of the hardware graph.

We showed that the designed gate set can afford a number of improvements over unoptimized pulses. Through DD demonstrations, the gate set is shown to substantially reduce the decay of multi-qubit states, achieving decay rates that are 30\% of the decay rates estimated for unoptimized Gaussian pulses for an eight-qubit system. Gate characterization via RB conveys that the gate set yields high-fidelity operations. Single-qubit gate error rates are boosted by 0.5\%, while simultaneous RB gate error rates can reach as high as 1.6\% percent. 

The final demonstration focused on Hamiltonian simulation. We showed that the optimized gate set can result in a median improvement in algorithm performance by a factor of 4 over unoptimized Gaussian pulses. To the best of our knowledge, this is the first hardware demonstration of optimized control (i.e. physical layer methods) leading to enhanced algorithm performance.

Lastly, through our analysis, we showed that co-design of CRGS and statically-coupled transmon architectures presents a significant opportunity. For example, these qubits occupy less space on-chip and require fewer control lines as compared to alternative transmon architectures with more tunability. This simplified design could unlock many opportunities for scaling qubit yields. Unfortunately, parasitic $ZZ$ crosstalk intrinsic to the fixed coupler design must be addressed if these devices are to stay competitive in the current quantum technology ecosystem. CRGS represent progress in overcoming this obstacle that is scalable and easily deployable on existing quantum systems. IBM cloud-based calibration limited our ability to demonstrate state-of-the-art performance of single qubit gates, but we do not expect this to be a fundamental limit, as demonstrated by our successful calibration at shorter durations on APL hardware. Incorporation of recent, improved crosstalk metrics into the optimal control framework can also help boost crosstalk suppression~\cite{kamen2026comparing}.

Future work would focus on extending gate design to two-qubit gates as well. The ECR entangling gate proved to be a limiting factor in our IBM experiments. In practice, two-qubit gate schemes should be designed explicitly for the strong-coupling regime, potentially with error-robust pulse shaping of a direct CNOT~\cite{long2021universal,malekakhlagh2022mitigating}. Other software approaches have been suggested for addressing two-qubit crosstalk throughout the ECR gate, and integration of these approaches may further improve the success of our gate sets~\cite{kim2023evidence,seif2024suppressing}. Another interesting line of effort would be to augment our gate set with DRAG cancellation tones for microwave crosstalk or leakage suppression~\cite{hyyppa2024reducing, wesdorp2026mitigating}, and to pursue a reduction of the gate set into an easy-to-calibrate, parametric form.

CRGS provide a scalable solution to the problem of crosstalk suppression across all quantum computing modalities. Moreover, our novel, coordinated optimal control problem demonstrates the value of thinking about optimal control strategies beyond individual gates and state preparations, in order to discover new ways for pulse engineering to improve the scalability of quantum processors.

\section*{Code availability}
\texttt{Piccolo.jl} is an open-source Julia package for quantum optimal control~\cite{piccolo2025}. The simulator used for pulse-level circuit dynamics is an open-source Python repository provided by the authors at~\cite{pulsesim2025}.

\section*{Acknowledgments} \label{sec:ackn}
The authors would like to thank Ali Javadi-Abhari for many helpful discussions during this work.
This work was supported in part by the U.S. Department of Energy, Office of Science, Office of Advanced Scientific Computing Research, Accelerated Research in Quantum Computing under Award Number DE-SC0020316 and DE-SC0025509,
and in part by the STAQ project under award NSF Phy-232580.
GQ acknowledges additional support from ARO MURI grant W911NF-18-1-0218. 
AJG was supported by an appointment to the Intelligence Community Postdoctoral Research Fellowship Program at University of Chicago administered by Oak Ridge Institute for Science and Education (ORISE) through an interagency agreement between the U.S. Department of Energy and the Office of the Director of National Intelligence (ODNI). 
This research used resources of the Oak Ridge Leadership Computing Facility, which is a DOE Office of Science User Facility supported under Contract DE-AC05-00OR22725.
We acknowledge the use of IBM Quantum services for this work. The views expressed are those of the authors, and do not reflect the official policy or position of IBM or the IBM Quantum team.

\appendix
\numberwithin{equation}{section}
\numberwithin{figure}{section}

\section{Robustness} 
Robust control can be achieved in a number of ways; in this work, we minimize the sensitivity of infidelity to a provided error Hamiltonian. We refer to this objective as the \textit{error susceptibility}. When applied to crosstalk errors, we use problem structure to reduce the size of the Hilbert space needed to compute the particular case of \textit{crosstalk susceptibility}.

\subsection{Infidelity sensitivity} \label{apdx:robustness}
Suppose we have a Hamiltonian, $H(t) = H_C(t) + H_N$, that includes a control, $H_C(t)$, and noise-generating Hamiltonian, $H_N$. It is convenient to investigate the dynamics of this system in the rotating frame with respect to the control dynamics $U_C(t)=\mathcal{T}_+ \exp\left(-i \int^{t}_0 H_C(s)ds\right)$, where $\mathcal{T}_+$ is the time-ordering operator. The resulting total evolution operator $U(t)=\mathcal{T}_+ \exp\left(-i \int^{t}_0 H(s)ds\right)$ can be expressed as $U(t)=U_C(t)\widetilde{U}_N(t)$, where $\widetilde{U}_N(t)=\mathcal{T}_+ \exp\left(-i \int^{t}_0 \widetilde{H}_N(s)ds\right)$ is the rotated-frame noise evolution. The dynamics of this evolution are generated by $\widetilde{H}_N(t)=U^{\dagger}_C(t)H_N U_C(t)$.

If the noiseless control dynamics are performed as expected, $U_C(T)=G$. Thus, the gate fidelity [see Sec.~\ref{sec:qc:smooth}] is given by
\begin{eqnarray}
   \mathcal{F}(\mathbf{Z}, G) &=& 1 - \frac{1}{d^2} |\Tr \{ G^\dagger U_C(T) \widetilde{U}_N(T)) \} |^2 \nonumber\\
   &=& 1 - \frac{1}{d^2} |\Tr{ \widetilde{U}_N(T) }|^2,
\end{eqnarray}
where $d$ is the dimension of the Hilbert space. This expression indicates that the efficacy of the control is determined by the protocol's ability to minimize the contribution of the noise to the overall dynamics. That is, ideally, the control-rotated noisy evolution is given by the identity operation: $\widetilde{U}_N(T)=I$.

Deviations from the ideal scenario are typically examined via time-dependent perturbation theory~\cite{green2013arbitrary,poggi2024universally,zhou2023quantum}. In the weak noise limit, $\|H_C(t)\| \gg \|H_N\|$ $\forall t$, where $\|A\|$ denotes a unitarily-invariant norm of $A$ (typically the operator norm). Noise contributions can be treated as a perturbation in this limit. Commonly, the perturbation theory is performed via the Magnus expansion, with
\begin{equation}
    \widetilde{U}_N(T) = \exp \left( - i \sum_{j=1}^\infty \lambda^j \widetilde{H}_N^{(j)}(T) \right),
\end{equation}
where we've added in a scale $\lambda$ to track orders of the perturbation, $H_N$. The relevant terms in this series are
\begin{align*}
    \widetilde{H}^{(1)}_N &= \frac{1}{T} \int_0^T dt \widetilde{H}_N(t) \\
    \widetilde{H}^{(2)}_N &= \frac{1}{2iT}\int_0^T dt_1 \int_0^{t_1} dt_2 [\widetilde{H}_N(t_1), \widetilde{H}_N(t_2)].
\end{align*}
Plugging this in, and noting that any Hamiltonian and commutator terms are traceless, we have
\begin{align*}
	&\frac{1}{d} \Tr{\widetilde{U}_N(T)} \\
	&= \frac{1}{d} \Tr \{ \exp\{ - i \sum_{j=1}^\infty \lambda^j \widetilde{H}_N^{(j)}(T) \} \} \\
	&= \frac{1}{d} \Tr \{ 
	1 
	- i \sum_{j=1}^2 \lambda^j \widetilde{H}_N^{(j)}
	+ \frac{(-i)^2}{2} \left(\lambda \widetilde{H}_N^{(1)} \right)^2
	+ \mathcal{O}(\lambda^3)
	\} \\ 
	&= 1 - \lambda^2 \frac{1}{2d} \Tr\{  \widetilde{H}_N^{(1)\, 2} \}
	+ \mathcal{O}(\lambda^3).
\end{align*}
To address the absolute value in the fidelity, use $|1-\lambda^2 a|^2 \approx 1 - 2 \lambda^2 \Re(a) + \mathcal{O}(\lambda^4)$. Putting it all together, the infidelity to second order in the perturbation is
\begin{equation*}
    1 - \frac{1}{d^2} |\Tr\{\widetilde{U}_N(T)\}|^2
    = \frac{\lambda^2 }{d} \Re\Tr\{ \widetilde{H}_N^{(1)}(T)^2 \}
    + \mathcal{O}(\lambda^3).
\end{equation*}
This is the robustness objective in the main text,
\begin{equation} \label{apdx:eqn:robustness}
    \mathcal{R}(\mathbf{Z}; H_N) = \Re \frac{1}{d} \Tr\{  \big(\frac{1}{T}  \sum\nolimits_{t=1}^T U_t^\dagger H_N U_t \big)^2 \},
\end{equation}
once the control evolution operator has been made piecewise-constant.

\subsection{Crosstalk sensitivity} \label{apdx:xtalk-robustness}
Given a graph of vertices $V$ and undirected edges $E$, we asserted in Eq.~\eqref{eqn:xtalk-factor} that the crosstalk robustness objective can be factored in the following way:
\begin{equation}
    \mathcal{R}\bigg(\mathbf{Z}; \sum_{i,j \in E} \sigma^{(i)}_A \sigma^{(j)}_B \bigg) = \sum_{i,j \in E} \bar{\mathcal{R}}\left(\mathbf{Z}_i, \mathbf{Z}_j; \sigma^{(i)}_A \sigma^{(j)}_B \right).
\end{equation}
Here, $\mathbf{Z}$ is a trajectory of optimization variables and $\sigma_A$ is the usual Pauli operator. For the crosstalk robust control problem, the control dynamics are independent; that is, $U_t = U_t^{(1)} U_t^{(2)} \cdots U_t^{(|V|)}$. Define $\widetilde{\sigma}_t := U_t^\dagger \sigma\, U_t$, and observe that $U_t^\dagger \sigma^{(i)}_A \sigma^{(j)}_B U_t = \widetilde{\sigma}_{A,t}^{(i)} \, \widetilde{\sigma}_{B,t}^{(j)}$ is local to each qubit. As such, Eq.~\eqref{apdx:eqn:robustness} becomes
\begin{align*}
    &\mathcal{R}(\mathbf{Z}; \sum\nolimits_{i,j \in E} \sigma^{(i)}_A \sigma^{(j)}_B) \\
    &= \Re \frac{1}{d} \Tr\{ \big(\frac{1}{T}  \sum\nolimits_{t=1}^T \sum\nolimits_{i,j \in E} \widetilde{\sigma}_{A,t}^{(i)} \,\, \widetilde{\sigma}_{B,t}^{(j)} \big)^2 \} \\
    &= \Re \frac{1}{d} \frac{1}{T^2} \sum_{t,t'=1}^T  \sum_{\substack{i,j \in E \\ i',j' \in E}} \Tr\{ \widetilde{\sigma}_{A,t}^{(i)} \,\, \widetilde{\sigma}_{B,t}^{(j)} \,\, \widetilde{\sigma}_{A,t'}^{(i')} \,\, \widetilde{\sigma}_{B,t'}^{(j')} \} \\
    \intertext{Note $\Tr\{\widetilde{\sigma}_{A,t}^{(i)}\} = 0$ for any unpaired operator, so}
    &= \Re \frac{1}{d} \frac{1}{T^2} \sum_{t,t'=1}^T  \sum_{\substack{i,j \in E \\ i',j' \in E}}  \delta_{i i'} \delta_{j j'} \Tr\{ \widetilde{\sigma}_{A,t}^{(i)} \,\, \widetilde{\sigma}_{B,t}^{(j)} \,\, \widetilde{\sigma}_{A,t'}^{(i')} \,\, \widetilde{\sigma}_{B,t'}^{(j')} \} \\
    &= \Re \frac{1}{d} \frac{1}{T^2}  \sum_{t,t'=1}^T \sum_{i,j \in E} \Tr\{ \widetilde{\sigma}_{A,t}^{(i)} \,\, \widetilde{\sigma}_{A,t'}^{(i)} \} \Tr\{ \widetilde{\sigma}_{B,t}^{(j)} \,\, \widetilde{\sigma}_{B,t'}^{(j)} \}.
\end{align*}
In the last two lines, we used the identity $\Tr{\mu\otimes \nu} = \Tr{\mu}\Tr{\nu}$ for any operators $\mu,\nu$.
Now, we can compute the crosstalk robustness condition using only the independent qubit trajectories,
\begin{align} \label{apdx:eqn:crosstalk-robustness}
    &\bar{\mathcal{R}} \left(\mathbf{Z}_i, \mathbf{Z}_j; \sigma^{(i)}_A \sigma^{(j)}_B \right)
    \\ \nonumber & := \Re \frac{1}{d} \frac{1}{T^2} \sum_{t,t'=1}^T \Tr\{ \widetilde{\sigma}_{A,t}^{(i)} \,\, \widetilde{\sigma}_{A,t'}^{(i)} \} \Tr\{ \widetilde{\sigma}_{B,t}^{(j)} \,\, \widetilde{\sigma}_{B,t'}^{(j)} \}
\end{align}

\subsection{ZZ crosstalk sensitivity} \label{apdx:analytic-xtalk-robustness}

From Eq.~\eqref{apdx:eqn:crosstalk-robustness},
\begin{align} \label{apdx:eqn:zz-crosstalk-robustness}
    &\mathcal{R}(\mathbf{Z}_1, \mathbf{Z}_2; \sigma^{(1)}_Z \sigma^{(2)}_Z) \nonumber \\
    &= \Re \frac{1}{d} \frac{1}{T^2} \sum_{t,t'=1}^T \Tr\{ \widetilde{\sigma}_{Z,t}^{(1)} \,\, \widetilde{\sigma}_{Z,t'}^{(1)} \} \Tr\{ \widetilde{\sigma}_{Z,t}^{(2)} \,\, \widetilde{\sigma}_{Z,t'}^{(2)} \}
\end{align}
is the specialization to our specific crosstalk form factor. We can use this specific form to confirm its phase invariance and to find an analytic, crosstalk-robust baseline.

\subsubsection{Phase invariance} 
In the main text (Sec.~\ref{sec:virtualz}), we showed that the $Z$ robustness for a single qubit is phase invariant. This result naturally extends to the two-qubit case and $ZZ$ robustness. W.l.o.g., we can assume a phase offset $\phi_2$ on the second qubit in Eq.~\eqref{apdx:eqn:zz-crosstalk-robustness}. As such, the phase dependence is entirely in $\Tr\{ \widetilde{\sigma}_{Z\,t}^{(2)} \,\, \widetilde{\sigma}_{Z\,t'}^{(2)} \}$. This is exactly the case discussed in the main text, so Eq.~\eqref{apdx:eqn:zz-crosstalk-robustness} is also phase invariant.

\subsubsection{Example: Offset gates}
We can calculate a crosstalk-free parallel $X$ operation using $2\pi$-offset $X$ gates implemented using square wave pulses. 
Let $U_t^{(j)} = \exp\{-i \theta_j t \sigma^{(j)}_X / 2 T\}$ for $j=1,2$, with $\theta_1 = \pi / 2$ and $\theta_2 = \pi / 2 + 2\pi$.
Note 
\begin{equation}
    \widetilde{\sigma}_{Z\,t} = e^{i \theta t \sigma_X/2 T} \sigma_Z e^{-i \theta t \sigma_X/2 T} = e^{i \theta t \sigma_X / T} \sigma_Z
\end{equation}
so the sensitivity in Eq.~\eqref{apdx:eqn:zz-crosstalk-robustness} becomes
\begin{align*}
    & \Re \frac{1}{d} \frac{1}{T^2} \sum_{t,t'=1}^T
    \Tr\{  e^{i \theta_1(t - t') \sigma^{(1)}_X / T} \}
    \Tr\{  e^{i \theta_2(t - t') \sigma^{(2)}_X / T} \} \\
    &= \Re \frac{1}{d} \frac{1}{T^2} \sum_{t,t'=1}^T \cos(\theta_1(t - t') / T) \cos(\theta_2(t - t') / T).
\end{align*}
Complete the analysis by replacing sums with integrals. For $\theta_1=\theta_2$, the integrand is even and remains positive, $\mathcal{R} > 0$. Offsetting the angles such that $\theta_2 - \theta_1 = 2\pi$, the functions are orthogonal, so $\mathcal{R} = 0$.

\section{Pulse-level simulation} \label{apdx:simulation}
We simulate our circuits using a custom-built simulator~\cite{pulsesim2025} and Qiskit Dynamics~\cite{puzzuoli2023qiskit}. We model the pulse-level dynamics of the circuit using a Lindblad master equation in order to capture the impact of decoherence~\cite{krantz2019quantum}. The Lindblad equation
\begin{equation}
    \dot{\rho} =  - i[H, \rho] + \mathcal{D}_1(\rho) + \mathcal{D}_\phi(\rho)
\end{equation}
includes a trace-zero Hamiltonian term, $H$, and dissipation generating terms $\mathcal{D}_1(\rho)$ and $\mathcal{D}_\phi(\rho)$. We model \textit{longitudinal relaxation}
\begin{equation}
    \mathcal{D}_1(\rho) = \frac{1}{T_1}\left(  \sigma_{-} \rho \sigma_{-}^\dagger - \frac{1}{2} \left\{\sigma_{-}^\dagger \sigma_{-}, \rho \right\} \right),
\end{equation}
where $T_1$ is the characteristic time and $\sigma_{-}$ is the qubit lowering operator. \textit{Pure dephasing} is modeled using the characteristic time $T_\phi$ and
\begin{equation}
    \mathcal{D}_\phi(\rho) = \frac{1}{2T_\phi} \left( \sigma_Z  \rho \sigma_Z ^\dagger - \frac{1}{2} \left\{\sigma_Z ^\dagger \sigma_Z , \rho \right\} \right).
\end{equation}
The familiar transverse relaxation, $T_2$---which measures relaxation of a superposition state---combines $T_1$ and $T_\phi$ such that $T_2 \le 2 T_1$~\cite{krantz2019quantum}.

\begin{table}[t]
    \centering
    \begin{tabular}{p{3.5cm}P{1.5cm}P{1.5cm}}
    \hline
    Parameter & Mean & Std. \\
    \hline
    $T_1$ $\mu{s}$ & $216.0$ & $8.0$ \\
    $T_2$ $\mu{s}$ & $154.0$ & $8.2$ \\
    Frequency, $\omega$ (GHz) & $4.90$ & $0.11$ \\
    Anharmonicity, $\alpha$ (GHz) & $-0.31$ & $0.0054$ \\
    Coupling, $J$ (GHz) & $0.00172$ & $0.0$ \\
    Detuning, $\Delta$ (GHz) & $0.0$ & $0.01$ \\
    \hline
    \end{tabular}
    \caption{
    Model parameters set by IBM Brisbane: mean and standard deviation (Std.). Time values are in microseconds ($\mu{s}$) and frequencies in gigahertz (GHz). Crosstalk is calculated using Eq.~\eqref{eqn:crosstalk}.
    }
    \label{tab:model_params}
\end{table}
The parameters of the simulator are taken from records of IBM Brisbane~\cite{abughanem2025ibm}. A summary of the relevant parameters are shown in Table~\ref{tab:model_params}. $T_1$ and $T_2$ are set to reflect the best-performing qubits, and their standard deviation from the device is suppressed by an order of magnitude. More detailed empirical models of IBM devices can be obtained through characterization~\cite{oda2024sparse}.

\begin{figure}[t]
    \centering
    \includegraphics[width=\columnwidth]{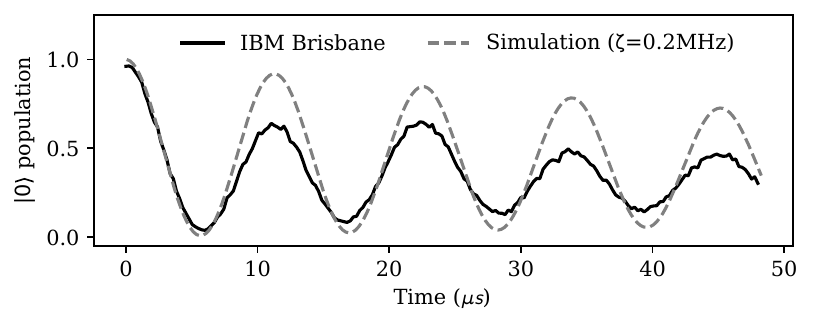}
    \caption{The ground state population of a default two-qubit $XY4$ experiment, and the corresponding result from the circuit simulator.}
    \label{fig:xy4-fit}
\end{figure}

The $ZZ$ crosstalk is determined by Eq.~\eqref{eqn:crosstalk}; it scales as $J^2$, and depends on a few device parameters, including the inter-qubit detuning. In our results, we do not model qubit layouts of the backends to capture this detuning. Instead, we universally set its value to the standard deviation of the qubit frequency distribution of the device. Our default simulation parameters are such that $ZZ$ is approximately $0.2$~MHz for all edges. As shown in Figure~\ref{fig:xy4-fit}, this reproduces the simultaneous $XY4$ experimental data from Sec.~\ref{sec:dd}. In addition to confirming our choice for $ZZ$, Fig.~\ref{fig:xy4-fit} also showcases our optimistic $T_1$ times (Table~\ref{tab:model_params}), as our simulation decays slower than the actual data.

For the two-qubit gates in our circuits, we simulate an \textit{echoed cross-resonance} (ECR) gate~\cite{sheldon2016procedure}. In the control-target Hilbert space (denote $1$ and $2$ for control and target, respectively), the ECR gate is the sequence
\begin{equation}
    U_\text{ECR}(\theta) := e^{-i \frac{\theta}{2} H_\text{CR}} \cdot \sigma^{(2)}_X \cdot e^{+i \frac{\theta}{2} H_\text{CR}}.
\end{equation}
where $H_\text{CR}$ is the Hamiltonian of the \textit{cross-resonance} (CR) drive---a microwave drive of a \textit{control} qubit at the resonance frequency of a \textit{target} qubit. The leading-order effective model for $H_\text{CR}$ is~\cite{sheldon2016procedure, magesan2020effective, malekakhlagh2022mitigating}:
\begin{align}
    H_\text{CR}(a) = a \cdot \frac{J}{\Delta_{12} + \alpha_1}\left( - \sigma^{(2)}_X + \frac{\alpha_1}{\Delta_{12}} \sigma^{(1)}_Z \sigma^{(2)}_X \right).
\end{align}
The ECR gate uses its \textit{echo}, $\sigma^{(2)}_X$, to refocus some errors and approximately achieve $U_\text{ECR}(\theta) \approx e^{-i \frac{\theta}{2} \sigma^{(1)}_Z \sigma^{(2)}_X}$. To cancel additional errors, a \textit{rotary} drive can be played on the target alongside the CR drive, $H_\text{CR}(a_1) + a_2 \sigma^{(2)}_X$~\cite{sheldon2016procedure, sundaresan2020reducing}. In our experiments, we simulate the full ECR sequence, driving with flat-top Gaussian envelopes for both a CR and rotary. Our rotary control is set to always counteract the single qubit term in the CR drive. To isolate the impact of our single-qubit moments, we also simulate our circuit using a two-qubit gate that is robust---meaning it suppresses $ZZ$ errors in the control, target and  target, target-spectator subspaces~\cite{watanabe2024zz}, while the echo cancels the $ZZ$ contribution of the control, control-spectator subspace~\cite{sundaresan2020reducing}. Our robust scheme is implemented by replacing the Gaussian envelope of the cross-resonance drive with a detuning-robust envelope---in particular, a scaled version of $\sqrt{X}$ from Sec.~\ref{sec:ibm-experiments} and Fig.~\ref{fig:gates}.

We simulate our circuit by decomposing it into one and two qubit moments. Taking advantage of the GPU simulation available in Qiskit Dynamics, we simulate all unique one qubit moments in parallel on GPU. We then do the same for the unique two qubit moments. Afterward, the full circuit is reassembled by multiplying unitaries.

\section{Rough and fine calibration} \label{apdx:calibration}
\begin{figure}[t]
    \centering
    \includegraphics[width=\columnwidth]{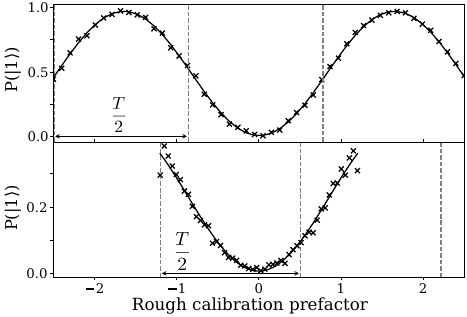}
    \caption{Successful (top) and unsuccessful (bottom) rough calibration of crosstalk-robust pulses on IBM Brisbane (2048 shots). The dotted vertical lines demarcate the half-period intervals of the calibration fits. The unsuccessful calibration is due to a persistent jump in the population, visible here upon its onset at the left and right endpoints of the bottom data. We scale our pulse durations by $4 \times$ to prevent such unsuccessful calibrations.}
    \label{fig:good-bad-calibration}
\end{figure}

The first step in our calibration pipeline is a Rabi experiment or \textit{rough calibration}, which scans a range of amplitude factors to find the one that best achieves our desired operation. For each amplitude, we estimate the $|1\rangle$ population from many experiments. Then, we fit the $|1\rangle$ population to a sinusoidal function, 
\begin{align}
    y = \frac{1}{2}\cos(2\pi \cdot x \cdot f + \pi) + \frac{1}{2}
\end{align}
where $x$ is the pulse amplitude and $y$ the $|1\rangle$ population. The fit parameter $f$ tells us the oscillation frequency of the given curve; hence, the calibrated amplitude for our pulse is
\begin{align}
    a_\textrm{rough} = \frac{\theta}{2\pi \cdot f},
\end{align}
where $\theta$ is the target angle of the gate. On IBM devices, the rough calibration curves would sometimes experience persistent jumps in population upon exceeding certain drive amplitudes, as if the drive units changed. Fig.~\ref{fig:good-bad-calibration} shows a successful and unsuccessful rough calibration, including the earliest onset of a jump (our amplitude sweep exits once the jump occurs).

After rough calibration, we refine our amplitude through the iterative \textit{fine calibration} defined in \cite{sheldon2016characterizing}. The goal of this step is to use error amplifying sequences to correct for a small, leftover, additive error $\Delta \theta$, which can be amplified by multiple repetitions of the pulse. We fit the population versus repetition data to a sinusoidal function to extract this error and correct. When calibrating $\pi$ pulses, the circuits start with a $\pi / 2$ pulse to ensure that the measured population is centered around the $|+\rangle$ state, and then our $\pi$ pulse is applied some number of times. When calibrating $\pi/2$ pulses, we use an odd number of our $\pi / 2$ pulse, which guarantees the population is always around the equator of the Bloch sphere.

Upon running the experiment for a sequence of length 14 for $X$~gates (25 for $\sqrt{X}$~gates), we fit the normalized population to a cosine function. For $\pi$ pulses, we fit to
\begin{align}
    y = \frac{1}{2} \cos((\Delta \theta + \pi) x - \frac{\pi}{2}) + \frac{1}{2}
\end{align}
where the $\pi/2$ phase comes from the $\pi/2$ pulse applied at the beginning of each experiment. For $\pi / 2$ pulses, we fit to 
\begin{align}
    y = \frac{1}{2} \cos(\left(\Delta \theta + \frac{\pi}{2}\right)x - \pi) + \frac{1}{2}.
\end{align}
Given the estimated $\Delta \theta$, we update our amplitude as 
\begin{align}
    a_\textrm{fine} = a_\textrm{rough} \cdot \frac{\theta}{\theta + \Delta \theta}.
\end{align}
This step can be repeated multiple times until $\Delta \theta$ is made smaller than some tolerance.

\section{APL Superconducting Qubit System}
\label{apdx:apl-system}
The in-house hardware used in this study consists of a fixed-frequency, fixed-coupler transmon-based superconducting qubit device with six qubits coupled in a ring topology. The qubit used for the experiments shown here possess a qubit frequency of $\approx 5.4$ GHz, with $T_1\approx 59\mu$s and $T_2\approx 38\mu$s. The device is housed in a $\mu$-metal shield under the mixing
chamber stage of a dilution refrigerator, held at 20 mK. A
readout resonator at $\approx$ 7 GHz is coupled to the qubit, and the state of the qubit is determined by standard cQED techniques.

%

\bibliographystyle{apsrev4-2}
\bibliography{bibs/crosstalk,bibs/arxiv}

\end{document}